\documentclass[aps,pre,twocolumn,groupedaddress,showpacs]{revtex4}
\usepackage{bm,graphicx}
\usepackage{amsmath}
\usepackage{amsfonts}
\usepackage{float}
\newcommand{\sgn}{\text{sgn}}
\begin{document}
%

%Title of paper
\title {Flexoelectricity and pattern formation in nematic liquid crystals}
%

%Authors
%
\author{Alexei Krekhov}
\email[]{alexei.krekhov@uni-bayreuth.de}
%\homepage[]{Your web page}
%\thanks{}
%\altaffiliation{}
%
\author{Werner Pesch}
\affiliation{Physikalisches Institut, Universit\"at Bayreuth,
D-95440 Bayreuth, Germany}
\author{\'Agnes Buka}
\affiliation{Research Institute for Solid State Physics and Optics,
Hungarian Academy of Sciences, H-1525 Budapest, P.O.B.49, Hungary}
\date{\today}
\begin{abstract}
We present in this paper a detailed analysis of the flexoelectric instability of a planar nematic layer in the presence of an alternating electric field (frequency $\omega$), which leads to stripe patterns (flexodomains) in the plane of the layer.
This equilibrium transition is governed by the free energy of the nematic which describes the elasticity with respects to the orientational degrees of freedom supplemented by an electric part.
Surprisingly the limit $\omega \to 0$ is highly singular.
In distinct contrast to the dc-case, where the patterns are stationary and time-independent, they appear at finite, small $\omega$ periodically in time as sudden bursts.
Flexodomains are in competition with the intensively studied electro-hydrodynamic instability in nematics, which presents a non-equilibrium dissipative transition.
It will be demonstrated that $\omega$ is a very convenient control parameter to tune between flexodomains and convection patterns, which are clearly distinguished by the orientation of their stripes.
\end{abstract}
%

% insert suggested PACS numbers in braces on next line
% insert suggested keywords - APS authors don't need to do this
%\keywords{}

%\maketitle must follow title, authors, abstract, \pacs, and
%\keywords
\maketitle

%%%
%
\section{Introduction}
\label{sec:intro}
Nematic liquid crystals (nematics) are materials which prefer in some temperature range a uniaxial mean orientational ordering of their non-spherical molecules, while the positional order is fluid-like.
The locally preferred axis is described by the director field $\bm{n}$ with $\bm{n}^2=1$ \cite{de Gennes, Chandrasekhar:1992, book}.
The basic equilibrium state, in the absence of external stresses, corresponds to a spatially uniform director configuration, where the orientational elastic free energy, $F_{el}$, with respect to $\bm n$ takes a minimum.
Thus spatial variations of $\bm{n}$, which can be decomposed into splay, twist and bend distortions with the elastic constants $k_{11}$, $k_{22}$, $k_{33}$, respectively, lead to an increase of $F_{el}$, i.e., to torques on $\bm{n}$ to restore equilibrium.
In the presence of an electric field, $\bm{E}$, and in non-equilibrium configurations electric torques on the director have to be considered as well.
They originate from a nonzero electric polarization, $\bm P$, which contains at first the standard dielectric contribution $\bm{P}_{diel} = \epsilon_0 (\bm{\epsilon}-1) \bm{E}$.
The dielectric tensor, $\bm{\epsilon}$, which depends on the local director orientation, is governed by the two dielectric permittivities $\epsilon_{\parallel}$ and $\epsilon_{\perp}$ for $\bm{E}$ parallel and perpendicular to $\bm{n}$, respectively; $\epsilon_0$ denotes the vacuum permittivity.
Furthermore, in a rough analogy to the piezo-electric polarization of certain insulators due to mechanical strains, director distortions lead also to the so called flexo polarization, $\bm P_{fl} =e_1\bm{n}(\nabla\cdot\bm{n}) + e_3(\bm{n}\cdot\nabla)\bm{n}$, characterized by the two flexo coefficients $e_1$, $e_3$.
If the electric field is strong enough the balance of electric and elastic torques may even require spatially periodic director variations where typically the flexoelectric torques play a crucial role.
The main goal of this paper is a detailed study of the resulting patterns, which are easily detected by optical means due to the accompanying variations of the refraction index.
Our analysis is restricted to the mostly used {\em planar} director configuration, which is very convenient for experimental and theoretical studies.
In this case a uniform basic state is achieved by sandwiching a thin nematic layer between two plates (parallel to the $x-y$ plane), which may also serve as electrodes for the application of a transverse electric field $\bm{E}$.
By a suitable treatment of the plate surfaces the same in-plane director orientation $\bm{n} = \bm{n}_0 = \bm{\hat{x}}$ is enforced at the confining plates, which then persists throughout the whole layer in the basic state.
Before concentrating on the impact of the flexo torque we will briefly touch on the familiar consequences of the dielectric torque.
It is destabilizing for a {\em positive} dielectric anisotropy, $\epsilon_a = \epsilon_{\parallel} - \epsilon_{\perp} > 0$, in the planar case.
Thus above a certain field amplitude the compensation of the dielectric torque by the stabilizing elastic one becomes impossible and the {\em Freedericksz transition} takes place: the director experiences a distortion in the $x-z$ plane, maximal at the midplane of the nematic layer \cite{de Gennes}.
The resulting director variation is in most cases uniform in the plane of the layer and varies only in the $z$-direction.
Only in some exceptional cases, when $k_{22}/k_{11} < 0.303$, the homogeneous Freedericksz transition for $\epsilon_a > 0$ is replaced by the so called splay-twist Freedericksz transition which leads to director distortions in the form of stripes parallel to $\bm{n}_0$.
We are only aware of one experimental realization \cite{Lonberg:1985}, since almost exclusively $0.5 < k_{22}/k_{11} < 1$ for nematics.
Just for completeness, we mention that during the temporal evolution of the homogeneous Freedericksz state {\em transient stripe patterns} parallel to the $x$-axis have also been described \cite{Buka:1992}.
We now turn to the effect of a finite flexo polarization, which provides a robust mechanism (independent on the sign of $\epsilon_a$) to generate stripe patterns ({\em flexodomains}) parallel to $\bm{n}_0$ as well.
The pattern forming instability of the basic state takes place when the flexo torque ($\propto |e_1-e_3|$) is sufficiently strong compared with the elastic and dielectric torques \cite{Bobylev:1977}.
Flexodomains are indeed observed in several nematic materials; also for $\epsilon_a < 0$, where a Freedericksz transition is excluded \cite{Barnik:1977, Barnik:1978, May:2008}.
In this paper we present a comprehensive analysis of the flexodomains in the presence of an electric field $\bm{E} = \bm{\hat{z}} E_0 \cos(\omega t)$, both in the dc-case (circular frequency $\omega =0$) and the ac-case for small $\omega$.
Unfortunately a previous analysis of the dc-case \cite{Hinov:2009, Marinov:2010} suffers from a serious error.
So far we have concentrated on {\em equilibrium transitions} driven by an electric field, which are governed by a minimization principle of the free energy.
In nematics, however, we find often under the term electro-hydrodynamic convection (EC) a {\em non-equilibrium, dissipative transition} from the basic state towards a periodic arrangement of convection rolls.
Since the resulting stripes run rather perpendicular to the initial director orientation $\bm{n}_0$ they are easily distinguishable from the flexodomains.
The mechanism of EC, which has been first elucidated by Carr and Helfrich \cite{Carr:1969, Helfrich:1969} in pioneering papers, has been comprehensively discussed in a number of reviews in the last years (see, e.g., \cite{Buka:2006} and references therein).
Some new aspects of the effect of flexoelectricity on EC will be given in section~\ref{sec:ec} of this paper.
An interesting feature of our system is the possibility to switch between an equilibrium pattern-forming instability (flexodomains) and the dissipative EC instability just by tuning the ac-frequency $\omega$.
This scenario has indeed been observed in recent experiments on certain nematics \cite{May:2008}: In the dc-case and at very small $\omega$ one finds flexodomains, before EC takes over at increasing $\omega$.
Systematic theoretical analyzes, however, are practically missing so far.
To provide those has been an important issue of our work.
In section~\ref{sec:basic_eqns} we will briefly comment on the mathematical background of our theoretical investigations.
Section~\ref{sec:pikin} is devoted to the flexodomains.
In section~\ref{sec:ec} we concentrate on EC at low frequencies in comparison with the transition to flexodomains.
It will become obvious that in particular the limit $\omega \to 0$ is far from trivial and requires a careful analysis.
The paper concludes with some final remarks.
%

%%%
%
\section{Mathematical background}
\label{sec:basic_eqns}
Our starting point is a nematic layer (parallel to the $x-y$ plane) of thickness $d$ under the action of the applied potential $U(t) = E_0 d\cos(\omega t)$ in the $z$-direction.
In the quiescent basic state the director field is homogeneous throughout the layer ($\bm{n} = \bm{n}_0 =\bm{\hat{x}}$) and flow is absent.
Pattern forming instabilities are associated with a distortion, $\delta \bm{n}$, of $\bm{n}_0$.
They lead often to flow ($\bm{v} \ne 0$) and to a perturbation, $\phi$, of the applied voltage $U(t)$.
Since both the planar director orientation at $z =\pm d/2$ and $U(t)$ are considered to be fixed, the conditions $\phi =0$ and $\delta \bm{n} =0$ have to be fulfilled at the confining plates ($z =\pm d/2$).
This applies also to $\bm{v}$ under the realistic assumption of no-slip at the boundaries.
The general theoretical framework to analyze electrically driven pattern forming instabilities in nematic liquid crystals is well established in terms of the standard nemato-hydrodynamic equations \cite{de Gennes, Chandrasekhar:1992, book}.
The electric field distribution is determined by the Maxwell equations in the electro-quasi-static approximation.
As material parameters we need the dielectric permittivities $\epsilon_{\parallel}$, $\epsilon_{\perp}$, the electric conductivities $\sigma_{\parallel}$, $\sigma_{\perp}$ (for $\bm{n}$ parallel and perpendicular to the electric field $\bm{E}$), and the flexo coefficients $e_1$, $e_3$.
The director dynamics is driven by elastic and electric torques; in the presence of flow also viscous torques have to be taken into account. 
The velocity field is determined by the (generalized) Navier-Stokes equation, where the viscous stress tensor depends on the orientation of $\bm{n}$ with respect to $\bm{v}$ and its gradients.
For explicit calculations one needs the values of the elastic constants $k_{ii}, i =1,2,3$ and furthermore the five independent viscosity coefficients $\alpha_i, i=1, \dots, 5$ to quantify the stress tensor and the viscous torques.
It is natural to introduce dimensionless material parameters of order one (labeled by primes).
They appear then in the non-dimensionalized basic equations (see, e.g., \cite{Krekhov:2008}) and are usually defined as follows:
\begin{eqnarray}
\label{eq:scale}
&& k_{ii} = k'_{ii} k_0 \;, \;
\alpha_i = \alpha'_i \alpha_0 \;, \;
\nonumber \\
&& (\sigma_{\parallel}, \sigma_\perp) = 
(\sigma'_{\parallel}, \sigma'_\perp) \sigma_0 \;,
\nonumber \\
&& (e_1, e_3) = (e'_1, e'_3) \sqrt{\epsilon_0 k_0} \;, \;
\end{eqnarray}
with
\begin{eqnarray}
\label{eq:units}
&& k_0 = 10^{-12}~\text{N} \;, \;
\alpha_0 = 10^{-3}~\text{Pa~s} \;, 
\nonumber \\
&& \sigma_0 = 10^{-8}~\text{($\Omega$~m)}^{-1} , \;
\epsilon_0 = 8.8542 \times 10^{-12}~\frac{\text{A~s}}{\text{V~m}} \;. \qquad
\end{eqnarray}
For quantitative calculations in this paper we refer to the standard nematic MBBA \cite{remark}, which has been used in many experimental investigations in the past.
The material parameters of MBBA are well known; for instance, for the analysis of flexodomains we use the following dimensionless elastic and electric constants (the primes are omitted):
\begin{eqnarray}
\label{eq:mat}
&& k_{11} = 6.66 \;, \;\; k_{22} = 4.2 \;,
\nonumber \\
&& e_1 = -3.25 \;, \;\; e_3 = -4.59 \;, \;\; 
\epsilon_a = -0.53 \;.
\end{eqnarray}
By using Eq.~(\ref{eq:scale}) it is conventional to measure lengths in units of $d/\pi$, time in units of $\tilde\tau$, and to introduce a dimensionless control parameter $R$ instead of the voltage amplitude $U_0$, where
\begin{eqnarray}
\label{eq:tau_r}
R = \frac{\epsilon_0 E_0^2 d^2}{k_0 \pi^2} = 
\frac{\epsilon_0 U_0^2}{k_0 \pi^2} \;, \;\;
\tilde \tau = \frac{\alpha_0 d^2}{k_0 \pi^2} \;.
\label{eq:tauR}
\end{eqnarray}
To describe the onset behavior of the pattern forming instabilities the nemato-hydrodynamic equations are linearized about the basic state.
As a result we arrive at a linear system of coupled partial differential equations in the variables $x$, $y$, $z$, $t$ for the perturbations $\delta \bm{n} = (0, n_y, n_z)$, $\bm{v}$, and $\phi$.
Since the lateral extensions of the nematic layer are much larger than the layer thickness $d$, periodic boundary conditions in the layer plane are appropriate. 
By switching accordingly with respect to the planar coordinates $\bm{x}=(x,y)$ to Fourier space $\bm{q}=(q,p)$, one gets the linear equations in the form to be found in the Appendix of \cite{Krekhov:2008}; a simplified version to analyze the flexodomains will be presented explicitly in section~\ref{sec:pikin}.
As already explained, the perturbations $\delta \bm{n}$, $\bm{v}$, $\phi$ are assumed to vanish at $z = \pm d /2$; this is guaranteed by expanding these fields in terms of a complete set of Galerkin trial functions which vanish at $z = \pm \pi/2$ in dimensionless units.
For instance the director component $n_z(\bm{q},z,t)$ is represented as:
\begin{eqnarray}
n_z(\bm{q},z,t) = \sum^{M}_{m=1}
\bar{n}_z(\bm{q},m,t) S_m(z) \;,
\label{eq:nzq}
\end{eqnarray}
with $S_m(z) = \sin[m (z +\pi/2)]$.
We have tested that using a truncation parameter $M =4$ yields already data to an accuracy of better than $0.1\%$; in many cases even $M =2$ is sufficient.
Introducing the symbolic vector $\bm{V}(\bm{q}, t)$ for the expansion coefficients $\bar{n}_z(\bm{q},m,t)$, $m=1,\ldots, M$ [Eq.~(\ref{eq:nzq})] and the corresponding ones for ${n}_y$, ${\bm{v}}$, ${\phi}$ we arrive after projection onto the trial functions at a linear system of coupled ordinary differential equations of the following form:
\begin{eqnarray}
{\cal C}(\bm{q},t) \frac{\partial}{\partial t} \bm{V}(\bm{q},t) = 
{\cal L} (R,\bm{q}, t) \bm{V}(\bm{q}, t) \;.
\label{eq:vsym}
\end{eqnarray}
The matrices ${\cal C}$ and ${\cal L}$ are periodic with the ac-voltage period $T=2\pi/\omega$.
The general solutions of Eq.~(\ref{eq:vsym}) have the Floquet representation $\bm{V}(t) = \exp(\sigma t) \bm{V}_0(t)$ with $\bm{V}_0(t+T) = \bm{V}_0(t)$, where $\sigma$ defines the Floquet exponent.
We are interested in time-periodic solutions $\bm{V}(t)$.
Here $\sigma$ has to be purely imaginary of the form $\sigma = i \omega k/l$ with co-prime integers $l > k$.
As a result the period of $\bm{V}(t)$ is given as $l \cdot T$ if $k \ne 0$ and as $T$ if $k =0$.
For a given $\bm{q}$ periodic solutions exist only for a discrete set of control parameters $R = R_0(\bm{q}) < R_1(\bm{q}) < R_2(\bm{q})$ etc.
Minimizing $R_0(\bm{q})$ with respect to $\bm{q}$ yields the critical wavevector $\bm{q}_c$ and the critical control parameter $R_c = R_0(\bm{q}_c)$ at which the quiescent basic state becomes unstable.
In the context of the present paper the destabilizing modes at onset had always the period $T$, i.e., they are characterized by $\sigma =0$.
We have used two methods to calculate the periodic solutions of Eq.~(\ref{eq:vsym}).
One option is to expand $\bm{V}_0(t)$ into a (truncated) Fourier series in terms of $\exp(i n \omega t)$ with $|n| \le N$.
Then Eq.~(\ref{eq:vsym}) transforms into an algebraic linear eigenvalue problem for the Fourier coefficients, from which we obtain $R_0(\bm{q})$ (for more details see \cite{Krekhov:2008}).
As will be documented below the time variations in $\bm{V}(t)$ become increasingly sharper with decreasing $\omega$; consequently many Fourier modes up to $N =60$ had to be eventually kept.
Thus, in an alternative, less time consuming approach, we construct numerically the matrix solution $\pmb{\mathcal P}(t)$ of Eq.~(\ref{eq:vsym}) for the initial condition $\pmb{\mathcal P}(0) = \bm{I}$ where $\bm{I}$ denotes the unit matrix (see, e.g., \cite{Hale} and for a recent application \cite{Heuer:2008}).
We have to calculate the eigenvalues $\mu_1, \mu_2, \dots$ of the ``monodromy matrix'' $\bm{M} \equiv \pmb{\mathcal P}(t =T)$ where $|\mu_1| > |\mu_2|$ etc. 
A periodic solution of Eq.~(\ref{eq:vsym}) with period $T$ exists when $\mu_1=1$.
The smallest $R$ to fulfill this condition yields again $R = R_0(\bm{q})$. 
As demonstrated in \cite{Krekhov:2008}, the general nemato-hydrodynamic equations as used in this paper, are invariant against a reflection $z \to -z$ at the midplane combined with a translation in time by half a period $T/2$.
Thus the solution manifold of Eq.~(\ref{eq:vsym}) naturally splits into separate classes with different parity $\pmb{\mathit p} = \pm 1$, for which the notions ``conductive'' ($\pmb{\mathit p} = 1$) and ``dielectric'' ($\pmb{\mathit p} = -1$), respectively, have been introduced.
For instance with respect to the director component ${n}_z$ one finds ${n}_z(-z,t+T/2) = {\pmb{\mathit p}} \, {n}_z(z,t)$ (for more details, see \cite{Krekhov:2008}).
This means that for the dielectric symmetry the time average of $n_z$ vanishes, while it is finite for the conductive symmetry.
%

%%%
%
\section{Flexodomains}
\label{sec:pikin}
In the following we will investigate the bifurcation to flexodomains which leads to stripe patterns with the wavevector $\bm{q} =(0,p)$.
It is easy to see that $U(t)$ is not modified, i.e., $\phi \equiv 0$.
Inspection of the full nemato-hydrodynamic equations (for instance in \cite{Krekhov:2008}), shows that time variations of the director at nonzero $\omega$ lead in principle to a ``back flow'' which acts back onto the director in the form of viscous torques.
They lead to small corrections to the dielectric and flexo torques of the order $O(\alpha_3^2/\eta^2_2)$, with the Miesowicz coefficient $\eta_2=(\alpha_3+\alpha_4+\alpha_6)/2 >0$.
Since $|\alpha_3/\eta_2|^2 = O(10^{-3} \div 10^{-4})$ for MBBA and similar nematics, the viscous torques are safely neglected in this paper, which also facilitates the quasi-analytical approaches in section~\ref{sec:acflexo} below.
Moreover, this approximation has been validated by full numerical studies of the basic equations.
It is convenient to introduce instead of the elastic constants $k_{11}$, $k_{22}$ their average value, $k_{av}$, and their relative deviation, $\delta k$, from $k_{av}$ as follows:
\begin{eqnarray}
\label{eq:dk}
 k_{11} = k_{av}(1 + \delta k) \;, \;\; 
 k_{22} = k_{av} (1- \delta k) \;,
\end{eqnarray}
where obviously $|\delta k| < 1$.
In contrast to the rod-like nematics like MBBA where $k_{22}<k_{11}$, i.e., $\delta k >0$ [see Eq.~(\ref{eq:mat})], discotic nematics are characterized by $k_{22}>k_{11}$ ($\delta k <0$) \cite{Stelzer:1997}.
Thus our analysis will cover negative $\delta k$ for completeness as well.
Neglecting the back flow effects ($\bm{v} =0$) the director dynamics in flexodomains is only determined by dielectric and flexo torques.
We start from the linear perturbations $\delta \bm{n}(y,z,t)$ of the basic state in position space and switch to Fourier space using the (real) ansatz:
\begin{eqnarray}
\label{eq:pikdeltn}
&& n_y(y,z,t) = \sin(p y) \bar{n}_y(z,t) \;,
\nonumber \\
&& n_z(y,z,t) = \cos(p y) \bar{n}_z(z,t) \;.
\end{eqnarray}
As a consequence the general equations (\ref{eq:vsym}) reduce to the following linear system of coupled partial differential equations (PDE's) in the variables $z$ and $t$ for the Fourier components $\bar{n}_y$ and $\bar{n}_z$:
\begin{subequations}
\label{eqn:nyz}
\begin{align}
 \partial_t \bar{n}_y &= 
-\left[ p^2 (1 + \delta k) - (1 - \delta k)\partial_{zz} \right] \bar{n}_y
 \nonumber \\
& + p \left[ \sgn(e_1 -e_3) u \cos(\omega t) 
   - 2 \delta k \partial_{z} \right] \bar{n}_z \;,
\label{eq:ny}\\
 \partial_t \bar{n}_z &= 
-\left[ p^2 (1 - \delta k) - \mu u^2 \cos^2(\omega t) 
- (1 + \delta k) \partial_{zz} \right] \bar{n}_z
\nonumber \\
& + p \left[ \sgn(e_1 -e_3) u \cos(\omega t) 
   + 2 \delta k \partial_{z} \right] \bar{n}_y \;.
\label{eq:nz}
\end{align}
\end{subequations}
Instead of the time scale $\tilde{\tau}$ and the main control parameter $R$ [Eq.~(\ref{eq:tau_r})] we have used in Eq.~(\ref{eqn:nyz}) the director relaxation time $\tau_d$ and the dimensionless voltage amplitude $u$, which are defined as follows:
\begin{eqnarray}
\label{eq:mu}
&& u^2 = \frac{1}{\mu} \frac{\epsilon_a}{k_{av}} R = 
       \frac{1}{\mu} \frac{\epsilon_a}{k_{av}}
       \frac{\epsilon_0 E_0^2 d^2}{k_0 \pi^2} \;, \;\;
\tau_d = (\gamma_1/k_{av}) \tilde{\tau} \;
\nonumber\\
&& \textrm{with} \;\; \mu = \frac{\epsilon_a k_{av}}{(e_1 -e_3)^2} \;\; \textrm{and} \;\; \gamma_1 = \alpha_3 - \alpha_2\;.
\end{eqnarray}
Note that we will often refer in the following to the parameter $\mu$ as a reduced measure of the dielectric anisotropy $\epsilon_a$.
Furthermore it should be realized that in Eqs.~(\ref{eqn:nyz}), (\ref{eq:mu}) only the difference of the flexo coefficients $(e_1-e_3)$ comes into play.
It is sufficient to confine oneself to the case $(e_1-e_3)>0$, since the solutions of Eqs.~(\ref{eqn:nyz}) for $(e_1-e_3) \to -(e_1-e_3)$ can be recovered by the transformation $\{ \bar{n}_y(z), \bar{n}_z(z) \} \to \{ \bar{n}_y(z), \bar{n}_z(-z) \}$. 
%

%%%
%
\subsection{Flexodomains driven by a dc-voltage}
\label{sec:dcflexo}
This subsection is devoted to the analysis of the flexodomains in the dc-case ($\omega=0$) where Eqs.~(\ref{eqn:nyz}) are exact due to $\bm{v} \equiv 0$.
Because all coefficients in Eqs.~(\ref{eqn:nyz}) are constant, the familiar separation ansatz $\bar{n}_{(y,z)}(z,t) = e^{\sigma t} \tilde{n}_{(y,z)}(z)$ can be used to get rid of the time dependence.
Thus the time derivatives $\partial_t$ in Eqs.~(\ref{eqn:nyz}) are replaced by $\sigma$ and one arrives at an autonomous linear system of ordinary differential equations (ODE's) for the functions $\tilde{n}_{(y,z)}(z)$.
As they have to vanish at $z = \pm \pi/2$, the eigenvalues $\sigma$ will belong to a discrete set.
The condition that the maximal eigenvalue $\sigma = \sigma_0(u,p)$ (growth rate) vanishes, determines the non-dimensional neutral curve $u_0(p)$.
The minimum of $u_0(p)$ at $p=p_c$ yields the critical voltage $u_c=u_0(p_c)$, where $(u_c, p_c)$ depend on $\delta k$ and $\mu$.
Alternatively $u_0(p)$ is given as the smallest $u$ value, where Eqs.~(\ref{eqn:nyz}) have time-independent solutions, which vanish at $z = \pm \pi/2$.
By the way, inspection of Eqs.~(\ref{eqn:nyz}) shows that $u_0(p)$ is even in $p$ since it depends only on $p^2$.
Let us first concentrate on the case of a destabilizing dielectric torque with positive $\epsilon_a$ where $\mu \propto \epsilon_a > 0$ [see Eq.~(\ref{eq:mu})].
For $p=0$ Eqs.~(\ref{eqn:nyz}) can be easily solved by choosing $\tilde{n}_y \equiv 0$ and $\tilde{n}_z \propto \sin(z + \pi/2)$.
As a result we obtain
\begin{eqnarray}
\label{eq:uF}
u_c(p =0)^2 \equiv u_F^2 = (1 + \delta k) /\mu \;.
\end{eqnarray}
This solution obviously describes the homogeneous splay Freedericksz distortion of the director.
In fact, we recover from $u_F^2$ with the help of Eq.~(\ref{eq:mu}) the familiar critical Freedericksz voltage $U_0 = U_F = \pi \sqrt{k_0 k_{av} (1 + \delta k)/(\epsilon_a \epsilon_0)}$ in physical units (which is not influenced by the flexo effect).
Clearly the condition $u_0(p_c) = u_c < u_F$ is necessary for the prevalence of flexodomains with wavenumber $p_c \ne 0$ against the homogeneous Freedericksz distortion.
It will turn out that they exist only for $\mu$ less than an upper limit $\mu_{max}(\delta k)$, at which their critical wavenumber $p_c$ approaches zero.
In the case of negative $\epsilon_a$ ($\mu < 0$), on the other hand, the dielectric torque is stabilizing.
It overcomes eventually the destabilizing flexo torques when $\mu$ approaches a lower limit $\mu_{min}(\delta k) < 0 $ from above, where $p_c $ diverges.
Thus the director remains undistorted in the basic planar state for $\mu < \mu_{min}(\delta k)$. 
In general, the $z$-dependence of the functions $\tilde{n}_{y,z}(z)$, which have to fulfill the ODE's introduced above, is captured by an ansatz $\propto e^{\lambda z}$.
In our case we obtain four different values $\lambda = \pm \lambda_1$, $\pm i \lambda_2$, where only real $\lambda_i$ are compatible with the existence of flexodomains.
The general solution, which consists of a linear combination of the four exponentials $e^{\pm \lambda_1 z}$, $e^{\pm i \lambda_2 z}$ has to fulfill the boundary conditions of vanishing $\tilde{n}_y$, $\tilde{n}_z$ at $z = \pm \pi/2$.
As shown in the Appendix, one arrive thus at the following implicit equation for the neutral curve $u_0(p)$:
\begin{eqnarray}
\label{eq:bc_det}
&& A_1 \sinh(\lambda_1 \pi) \sin(\lambda_2 \pi) 
\nonumber \\
&&\qquad + A_2 \lambda_1 \lambda_2 
 \left[ 1 - \cosh(\lambda_1 \pi) \cos(\lambda_2 \pi) \right] = 0 \;.
\end{eqnarray}
For the explicit expressions of the $\lambda_i$, $A_i$, $i=1,2$, which depend on $p$, $u$, $\delta k$, $\mu$, we refer to the Appendix as well. 
Equation~(\ref{eq:bc_det}) represents the neutral curve $u_0(p)$ in implicit form.
Minimization of $u_0(p)$ with respect to $p$ gives the critical wavenumber $p_c$ and the corresponding critical voltage $u_c \equiv u_0(p_c)$ of the flexodomains. 
%

%%% Figure 1
%
\begin{figure}[ht]
\centering
\includegraphics[width=7.0cm]{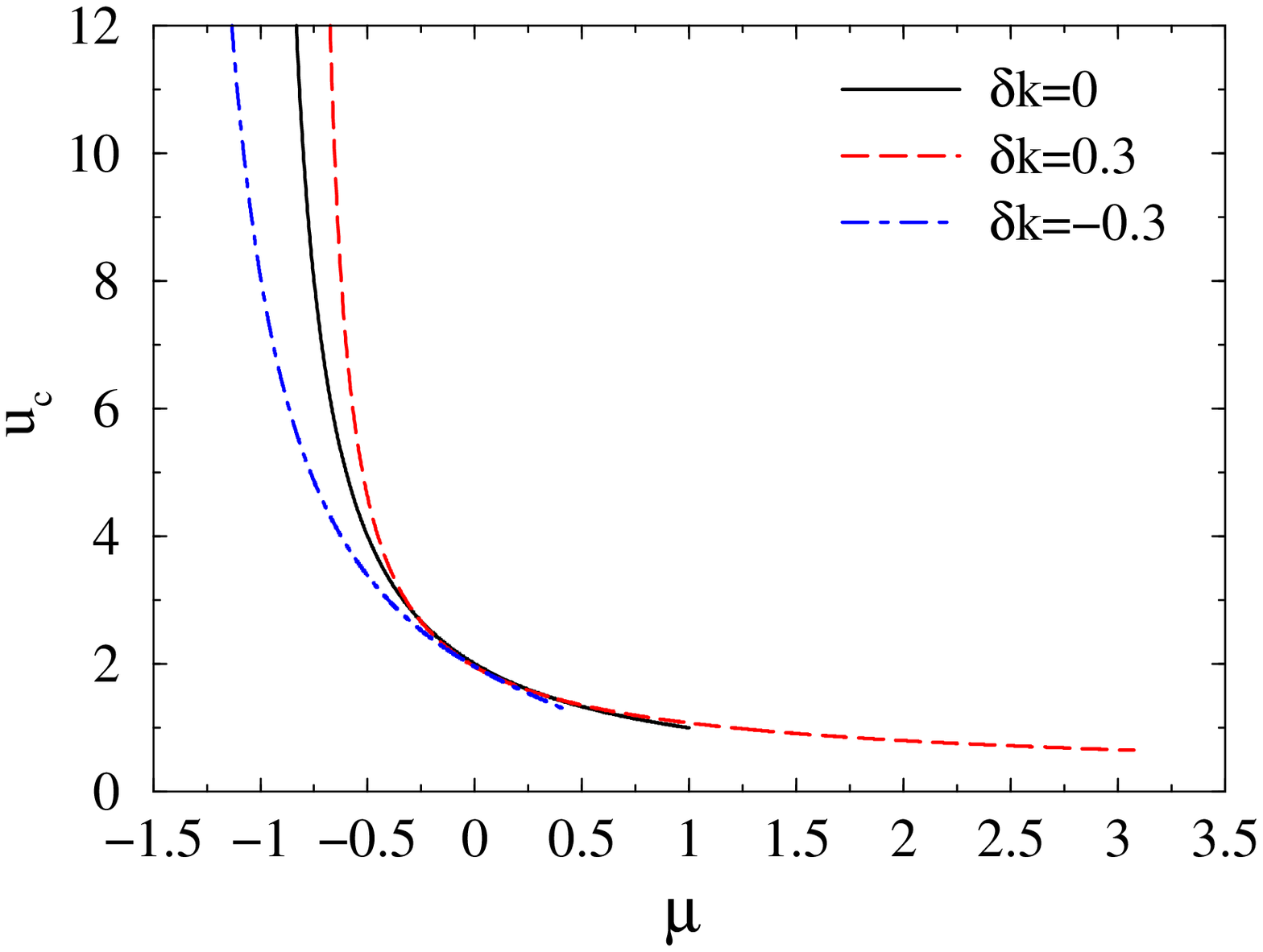}(a)
\includegraphics[width=7.0cm]{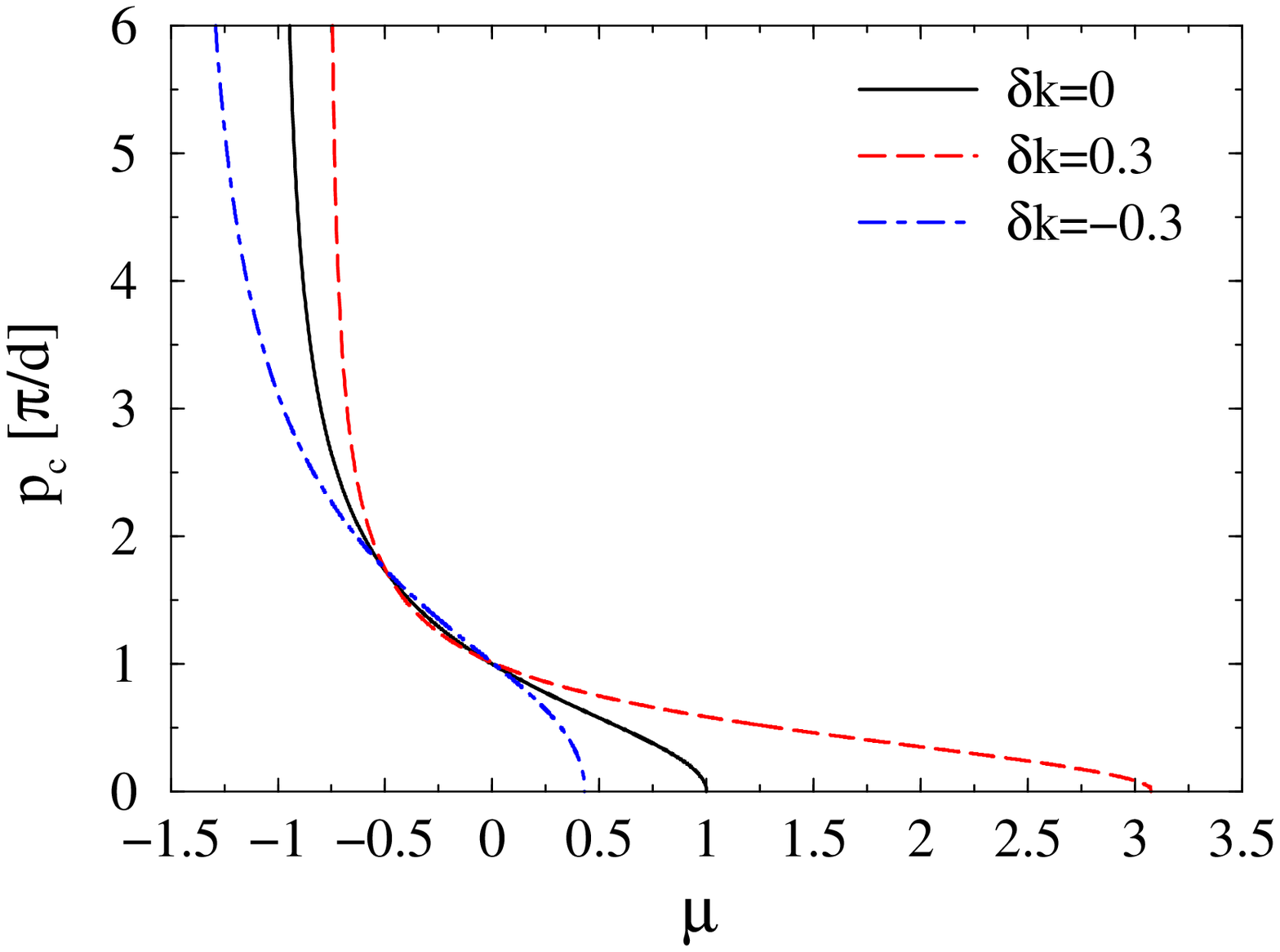}(b)
\caption{(Color online) Critical voltage $u_c$ (a) and critical wavenumber $p_c$ (b) as a function of $\mu$ for $\delta k =0$ ($\mu_{min} = -1$, $\mu_{max} = 1$), for $\delta k =0.3$ ($\mu_{min} = -0.769$, $\mu_{max} = 3.072$) and for $\delta k =-0.3$ ($\mu_{min} = -1.427$, $\mu_{max} = 0.413$).} 
\label{fig:uc_pc_mu_dc}
\end{figure}
\begin{figure}[ht]
\centering
\includegraphics[width=7.0cm]{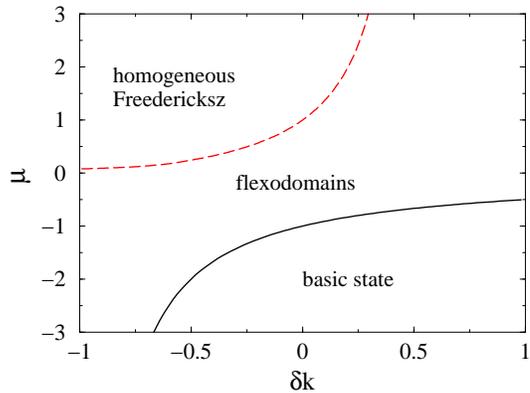}
\caption{(Color online) Upper and lower limit curves, $\mu_{max}(\delta k)$ (dashed, red) and $\mu_{min}(\delta k)$ (solid, black), respectively, in the $(\delta k - \mu)$ plane. $\mu_{max}(\delta k)$ diverges for $\delta k \approx 0.53$, while $\mu_{min} \to -\infty$ for $\delta k \to -1$.}
\label{fig:mu_min_max}
\end{figure}

In Fig.~\ref{fig:uc_pc_mu_dc}(a) we show representative examples for $u_c$ as a function of $\mu$ for $\delta k =0$ and for $\delta k =\pm 0.3$ calculated with the help of Eq.~(\ref{eq:bc_det}).
The corresponding critical wavenumbers $p_c$ are shown in Fig.~\ref{fig:uc_pc_mu_dc}(b).
As discussed before the Freedericksz state with $p_c =0$ and $u_c(0) = u_F$ is smoothly approached when $\mu \to \mu_{max}(\delta k)$.
Decreasing $\mu$ from $\mu_{max}$ on is associated with a monotonic increase of both $u_c$ and $p_c$ until they diverge at $\mu = \mu_{min}(\delta k) < 0$.
It is obvious that the knowledge of the limit curves $\mu_{min}(\delta k)$ and $\mu_{max}(\delta k)$ plays an important role to identify the regime of flexodomains in dependence on the parameters $\mu$ and $\delta k$.
Thus we show these limit curves in Fig.~\ref{fig:mu_min_max} in the $(\mu, \delta k)$ plane, where $|\delta k| <1$ according to Eq.~(\ref{eq:dk}).
On a first look, it is surprising that $\mu_{max}(\delta k)$ diverges at $\delta k \approx 0.53$.
However, as discussed in detail in the following subsection, this divergence is closely related to the existence of the (spatially periodic) splay-twist Freedericksz transition for $\delta k \gtrsim 0.53$ in the absence of the flexo torque ($e_1-e_3 =0$) \cite{Lonberg:1985}.
Before we turn, however, to further discussions of Eq.~(\ref{eq:bc_det}), we present at first a very useful approximate analysis of flexodomains for small $|\delta k|$. 
Instead of using Eq.~(\ref{eq:bc_det}) directly it is more transparent to start again from Eqs.~(\ref{eqn:nyz}), by introducing the ``one-mode'' approximation $\tilde{n}_{(y,z)} \propto \sin(z +\pi/2)$.
One obtains then immediately at a quadratic equation for the eigenvalues $\sigma$, the largest one determines the growth rate $\sigma_0$.
Note that in this approximation the terms $\propto 2 p \delta k \partial_z \tilde{n}_{(y,z)}$ in Eqs.~(\ref{eqn:nyz}) do not contribute.
Standard perturbation analysis shows that they would produce corrections of the order ${\delta k}^2$ to $\sigma_0$.
The condition $\sigma_0 =0$ leads within the one-mode approximation to the following expression for the neutral curve:
\begin{eqnarray}
\label{eq:r0dk}
 u^2_0(p) = \frac{(p^2+1)^2 - {\delta k}^2(p^2-1)^2}
 {p^2 + \mu[p^2+ 1 + {\delta k}(p^2 - 1)]}
\end{eqnarray}
with its minimum at $p^2=p_c^2$, where
\begin{eqnarray}
\label{eq:pcdk}
 p_c^2 = \frac{(-1 + {\delta k}^2) \mu + 
 \sqrt{(1 + {\delta k})[1 + {\delta k}(1 + 4 \mu)]}}
 {(1 + {\delta k})[1 + \mu(1 + {\delta k})]} \;. 
\end{eqnarray}
The explicit expression for $u_c^2=u^2_0(p_c)$ obtained from Eqs.~(\ref{eq:r0dk}), (\ref{eq:pcdk}) is quite lengthy and will not be shown.
According to the general remarks above, the $\mu$ interval, where flexodomains exist, can in general be read off from $p_c^2$ given in Eq.~(\ref{eq:pcdk}): the zero of the numerator determines the upper limit $\mu_{max}$ and the zero of the denominator the lower limit $\mu_{min}$.
Thus we obtain the following approximate expressions valid for small $\delta k$:
\begin{eqnarray}
\label{eq:lim_mu}
 \mu^a_{min}(\delta k) = -\frac{1}{1 + \delta k} \;, \;\; 
 \mu^a_{max}(\delta k) = \frac{1 + \delta k}{(1 - \delta k)^2} \;.
\end{eqnarray}
The one-mode approximation becomes exact in the special case $\delta k =0$ (one constant approximation, $k_{11}=k_{22}$), where it yields a rigorous solution of Eqs.~(\ref{eqn:nyz}).
We recover in this case the results of \cite{Bobylev:1977}:
\begin{eqnarray}
\label{eq:pcpik}
 p_c^2 = \frac{1-\mu}{1+\mu} \;, \;\;
 u_c^2 = \frac{4}{(1+\mu)^2} \;,
\end{eqnarray}
where $\mu_{min}(0) = -1$ and $\mu_{max}(0) = 1$.
In general the exact neutral curve from Eq.~(\ref{eq:bc_det}) is described to an accuracy of better than $0.5\%$ by Eq.~(\ref{eq:r0dk}) for small $|\delta k| < 0.2$.
We will demonstrate below, that $\mu^a_{min}(\delta k)$ given in Eq.~(\ref{eq:lim_mu}) even coincides with the exact curve $\mu_{min}(\delta k)$ shown in Fig.~(\ref{fig:mu_min_max}) for all $|\delta k| < 1$.
%

%%% Figure 2
%
Finally we would like to stress that our analysis of flexodomains is at variance with recent investigations \cite{Hinov:2009, Marinov:2010} on the same subject.
It will be explained in more detail in the Appendix that this work suffers from a basic mathematical error.
Thus for instance the prediction of a ``singular'' behavior of $p_c$ and $u_c$ both in the cases $k_{11}/k_{22} = 3$ ($\delta k = 1/2$) and $k_{11}/k_{22} = 1/3$ ($\delta k = -1/2$) does not hold.
%

%%%
%
\subsubsection*{Analytical treatment of $\mu_{max} (\delta k)$ and $\mu_{min} (\delta k)$}
\begin{figure}[ht]
\centering
\includegraphics[width=7.0cm]{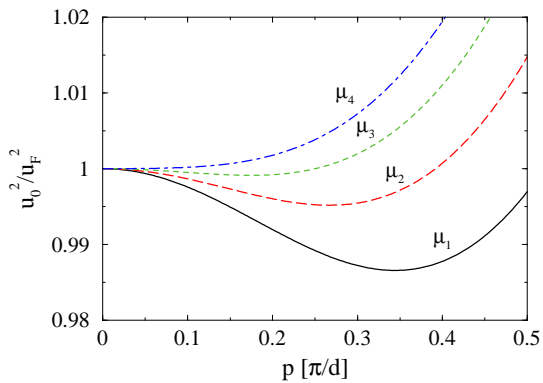}
\caption{(Color online) The neutral curve $u^2_0(p)$ normalized to the Freedericksz threshold $u_F$ (\ref{eq:uF}) as a function of $p$ for $\delta k = 0.2$ and different $\mu$ with $\mu_1 = 1.4, \mu_2 = 1.6, \mu_3 = 1.8, \mu_4 = 2.0$.}
\label{fig:neut_02}
\end{figure}

So far we have given the exact description of flexodomains through Eq.~(\ref{eq:bc_det}).
In addition we have demonstrated the usefulness of the analytical one-mode approximation (\ref{eq:r0dk}) of the neutral curve $u_0(p)$ at small $\delta k$.
In this section we will derive analytical expressions for the limiting curves $\mu_{min}(\delta k)$ and $\mu_{max}(\delta k)$ in the whole range $-1 < \delta k < 1$.
Let us start with the discussion of $\mu_{max}(\delta k)$ for the case $\mu > 0$, where we have competition between the flexodomains and the homogeneous Freedericksz state.
In Fig.~\ref{fig:neut_02} we show a typical neutral curve $u_0(p)$ for $\delta k = 0.2$ and different $\mu$ obtained from Eq.~(\ref{eq:bc_det}).
At $p=0$ the function $u_0(p)$ has an extremum with $u_0(0) = u_F$, see Eq.~(\ref{eq:uF}).
For $\mu < \mu_{max}(\delta k)$ this point corresponds to a maximum where $\partial^2_{p} u_0(p=0) < 0$; here and in the following the notation $\partial^n_p$ for the derivatives $(d^n/{d p^n})$ has been used.
The minimum of $u_0(p)$ at finite $p=p_c$ where $\partial^2_{p} u_0(p=p_c) > 0$ and $u_0(p_c) = u_c < u_F$ describes the flexodomains with wavenumber $p=p_c$.
%

%%% Figure 3
%
\begin{figure}[ht]
\centering
\includegraphics[width=7.0cm]{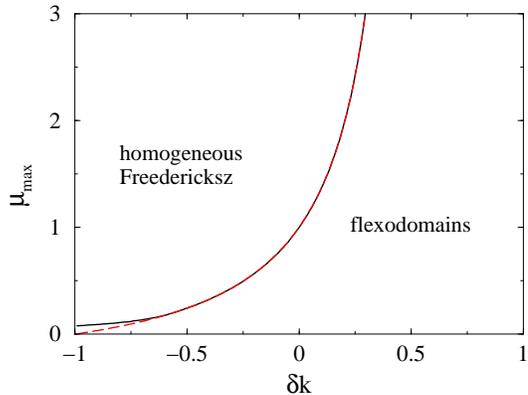}
\caption{(Color online) Upper limit curve $\mu_{max}$ (solid line, black) as a function $\delta k$ from Eq.~(\ref{eq:bc_det}) in comparison with $\overline{\mu}_{max}$ from Eq.~(\ref{eq_mu_up}) (dashed line, red).}
\label{fig:mu_upp}
\end{figure}

With increasing $\mu$ both the critical wavenumber $p_c$ and the difference $(u_F - u_c)$ decrease in Fig.~\ref{fig:neut_02}.
At $\mu = \mu_{max}(\delta k) = 1.965$ the minimum and the maximum of $u_0(p)$ merge at $p =0$.
Thus the equations $u_0(p=0) = u_F$ and $\partial^2 _{p} u_0(p =0)=0$ are fulfilled, which can be solved with respect to $\mu$ by expanding Eq.~(\ref{eq:bc_det}) up to order $O(p^2)$.
The resulting analytical solution is given as $ \mu = \overline{\mu}_{max}(\delta k)$ with
\begin{eqnarray}
\label{eq_mu_up}
 \overline{\mu}_{max}(\delta k) = 
 \frac{1 + {\delta k}}
 {1 - 2 {\delta k} + {\delta k}^2(32/\pi^2 - 3)} \;.
\end{eqnarray}
It is convenient to introduce also the function $\delta k_F(\mu)$ as the inverse of $\overline{\mu}_{max}(\delta k)$, which is given as: 
\begin{equation}
\label{eq:deltlow}
 {\delta k}_F(\mu) = 
 \frac{1 + 2 \mu  
 - \sqrt{(4\mu - 1)^2 - 128\mu(\mu-1)/\pi^2}}
 {2 \mu (32/\pi^2 - 3)}.
\end{equation}
Thus $\delta k_F(\mu)$ marks at fixed $\mu$ the transition from the homogeneous Freedericksz state for $\delta k < \delta k_F(\mu)$ to the flexodomains in the interval $\delta k_F(\mu)< \delta k < 1$ (see also Fig.~\ref{fig:mu_min_max}).
The special case ${\delta k_F(\mu = 1) =0}$ is consistent with $\mu_{max}(0) =1$ [see Eqs.~(\ref{eq:pcpik})].
From our reasoning it seems obvious that the analytical expression $\overline{\mu}_{max}(\delta k)$ should reproduce the curve $\mu_{max}(\delta k)$ shown in Fig.~\ref{fig:mu_min_max}.
As demonstrated in Fig.~\ref{fig:mu_upp} this is indeed the case for the interval $-{\delta k}_{low} < \delta k < 1$ where ${\delta k}_{low} \approx -0.566$.
On this $\delta k$ interval we will first concentrate.
%

%%% Figure 4
%
\begin{figure}[ht]
\centering
\includegraphics[width=7.0cm]{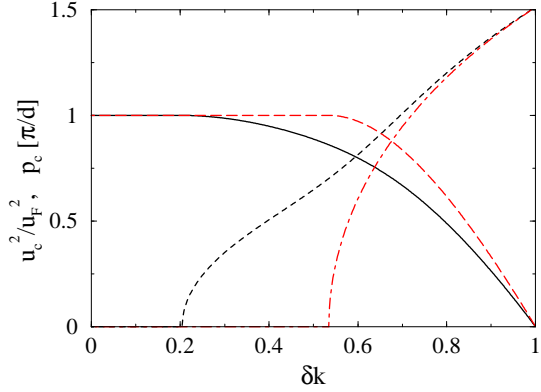}
\caption{(Color online) Relative critical voltage $u_c^2/u_F^2$ and the critical wavenumber $p_c$ as function of $\delta k$ for $\mu=2$ (solid line for $u_c^2/u_F^2$ and dotted line for $p_c$, black) and in the absence of flexo effect ($e_1-e_3=0$) (dashed line for $u_c^2/u_F^2$ and dot-dashed line for $p_c$, red).
For $\mu=2$ the Freedericksz state is approached at ${\delta k}_F \approx 0.206$, while this happens for $e_1-e_3=0$ at ${\delta k} = \delta k_{ST}$
[data from Eq.~(\ref{eq:bc_det})].}
\label{fig:ucpc_dk}
\end{figure}

One sees immediately that the denominator of $\overline{\mu}_{max}(\delta k)$ approaches zero (i.e., $\overline{\mu}_{max}$ diverges) when $\delta k$ approaches a critical value ${\delta k}_c > 0$ from below, where ${\delta k}_c$ is given as:
\begin{eqnarray}
\label{eq:deltkc}
 {\delta k}_c = \frac{1 - 2 \sqrt{1 - 8/\pi^2}}{32/\pi^2 - 3}
= 0.5346 \;.
\end{eqnarray}
Consistently $\delta k_F(\mu)$ in Eq.~(\ref{eq:deltlow}) approaches for $\mu \to \infty$ the limit $\delta k_c$.
As a result, flexodomains are exist for any $\mu >0$ when $\delta k > {\delta k}_c$, while they are restricted to the region $\mu < \overline{\mu}_{max}(\delta k)$ in the case of $\delta k < {\delta k}_c$.
It is not a coincidence that $\delta k_c$ is very near to the number $\delta k_{ST} \approx 0.53$, which has been quoted in the literature more than two decades ago in a different context: According to \cite{Lonberg:1985} the homogeneous Freedericksz transition, in the {\em absence of flexo effects} ($e_1-e_3=0$), is replaced for ${\delta k} > \delta k_{ST}$ by the spatially periodic splay-twist (ST) Freedericksz transition.
Similar to the flexodomains a state of finite $n_y, n_z$ bifurcates from the basic state which is periodic in the $y$-direction with a critical wavenumber $p_c$, to balance the dielectric and the elastic torques.
In this case additional flexo torques should not play a crucial role and the striped director configuration should develop even at arbitrary small $|e_1-e_3|$, which corresponds to an arbitrary large $\mu =\epsilon_a k_{av}/(e_1-e_3)^2$.
In fact, as shown in the Appendix, the limit $e_1-e_3 =0$ is covered by Eq.~(\ref{eq:bc_det}) and one finds $\delta k_{ST} \equiv {\delta k}_c$.
Our theoretical considerations are confirmed by the representative numerical results for the critical voltage $u_c$ and the critical wavenumber $p_c$ of flexodomains, which are shown in Fig.~\ref{fig:ucpc_dk} as function of $\delta k < 1$ for two different $\mu$.
In the case $\mu =2$ the homogeneous Freedericksz state is replaced at $\delta k = \delta k_F(\mu = 2) \approx 0.206$ by the flexodomains; with increasing $\delta k$ the critical voltage $u_c$ monotonically decreases from $u_c =u_F$ on, while $p_c$ increases from $p =0$ on. 
For $\mu \to \infty$ (i.e., in the absence of flexo effects) with ${\delta k}_F \to \delta k_{ST}$ we get in the interval ${\delta k}_{ST} \lesssim \delta k < 1$ the splay-twist Freedericksz distortion as a special case of the flexo patterns.
%

%%% Figure 5
%
\begin{figure}[ht]
\centering
\includegraphics[width=7.0cm]{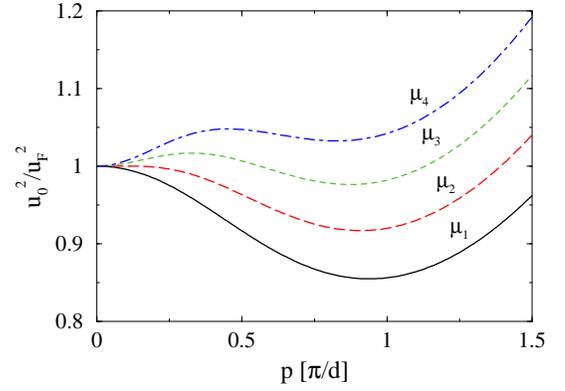}
\caption{(Color online) The neutral curve $u^2_0(p)$ as a function of $p$ for $\delta k = -0.7$ and different $\mu$ where $\mu_1 = 0.11, \mu_2 = 0.12, \mu_3 = 0.13, \mu_4 = 0.14$.}
\label{fig:neut_0_m7}
\end{figure}

We will now return to the case $-1 < \delta k < {\delta k}_{low} \approx -0.566$ in Fig.~\ref{fig:mu_upp}, where the exact $\mu_{max}(\delta k)$ from Eq.~(\ref{eq:bc_det}) is slightly larger than $\bar{\mu}_{max}(\delta k)$ given in Eq.~(\ref{eq_mu_up}).
The reasoning, which led to the expression $\bar{\mu}_{max}$, is rigorous as long 
as the $p$-dependence of neutral curves is of the type shown in Fig.~\ref{fig:neut_02}.
As demonstrated for instance in Fig.~\ref{fig:neut_0_m7} for $\delta k =-0.7$ this does not hold in the vicinity of $\delta k = -1$.
The condition $u_0(p)=u_F$, which determines $\mu_{max}(\delta k)$, is now already realized at a finite $p = p_c$, where $u_0(p)$ has a second local minimum of height $u_0(p_c) = u_F$.
Since a merging of the two extrema is not required, as assumed in the calculation of $\bar{\mu}(\delta k)$ from Eq.~(\ref{eq_mu_up}), we have now $\mu_{max}(\delta k) > \overline{\mu}_{max}(\delta k)$ in agreement with Fig.~\ref{fig:mu_upp}.
The exact neutral curve for $\delta k < {\delta k}_{low}$ is only accessible by numerically solving Eq.~(\ref{eq:bc_det}).
The special value $\delta k = {\delta k}_{low}$, where $\mu_{max}$ starts to deviate from $\mu_{max}(\delta k)$, is obviously determined by the condition $\partial^2_{p} u_0(p=0) = \partial^4_{p} u_0(p=0) =0$ at $u=u_F$ and $\mu=\mu_{max}(\delta k)$.
With the use of Eq.~(\ref{eq:bc_det}) one obtains from these conditions the following closed analytical expression for ${\delta k}_{low}$:
\begin{eqnarray}
 \delta k_{low} &=& -\Big[
 \frac{\pi^2 (96 - 5 \pi^2 - 8 \sqrt{\pi^4 - 54 \pi^2 + 468})}
      {-13 \pi^4 + 832 \pi^2 - 6912}
 \Big]^{1/2} 
\nonumber \\
&=& -0.5666 \;,
\end{eqnarray}
which agrees perfectly with the results presented in Fig.~\ref{fig:mu_upp}.
%

%%% Figure 6
%
Finally we consider the case $\mu \propto \epsilon_a < 0$, where the dielectric torque is stabilizing and where the competing Freedericksz transition is absent.
As already mentioned we found $\mu_{min}(\delta k) \equiv \mu^a_{min}(\delta k)$ given in Eq.~(\ref{eq:lim_mu}).
This can be explained by the fact, that exactly at $\mu = \mu^a_{min}$ the purely imaginary roots $\pm i \lambda_2$ in Eq.~(\ref{eq:bc_det}) become real (see the Appendix).
As a consequence the trigonometric functions $\sin(\lambda_2 \pi)$ and $\cos(\lambda_2 \pi)$ change into hyperbolic ones; then Eq.~(\ref{eq:bc_det}) allows only for the trivial solution $u=0$, $p=0$.
%

%%%
%
\subsection{Flexodomains driven by an ac-voltage}
\label{sec:acflexo}
\begin{figure}[ht]
\centering
\includegraphics[width=7.0cm]{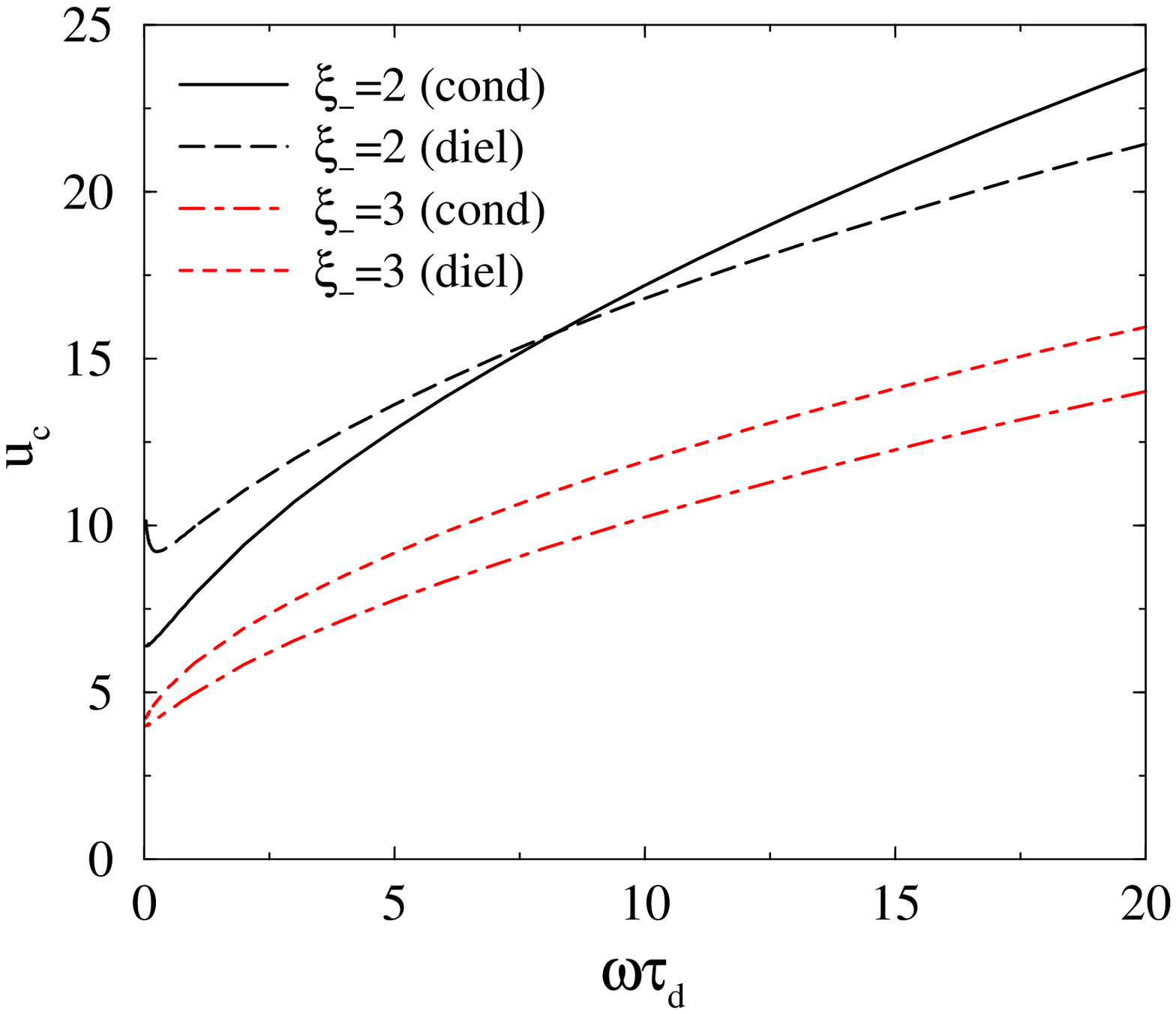}(a)
\includegraphics[width=7.0cm]{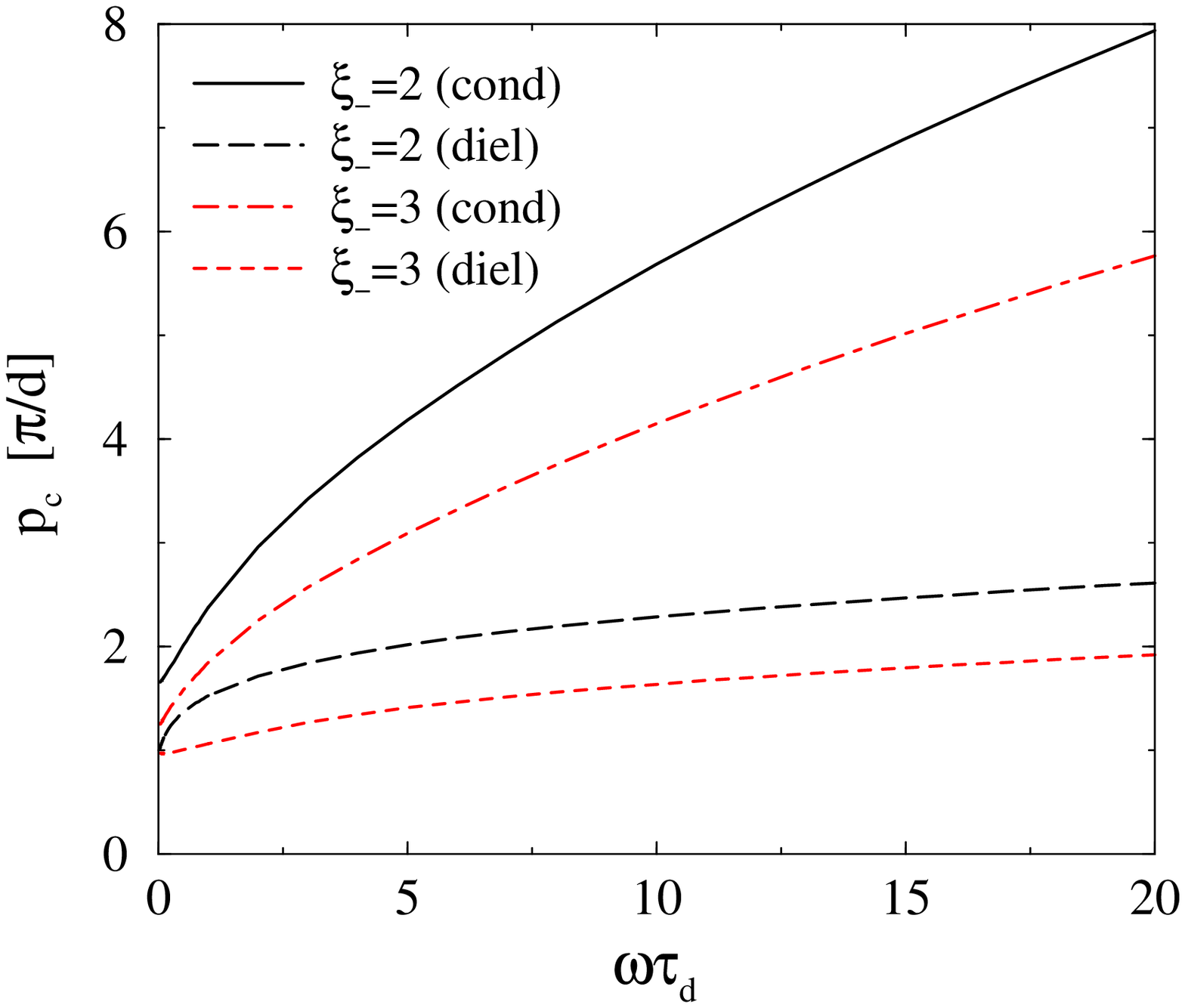}(b)
\caption{(Color online) Critical voltage $u_c(\omega)$ in dimensionless units (a) and critical wavenumber (b) of the flexodomains as function of $\omega \tau_d$ for MBBA parameters with $\xi_{-}=2$ and $\xi_{-}=3$.
The symmetry of the solution is indicated.}
\label{fig:uc_e13mod}
\end{figure}

\begin{figure}[ht]
\centering
\includegraphics[width=7.0cm]{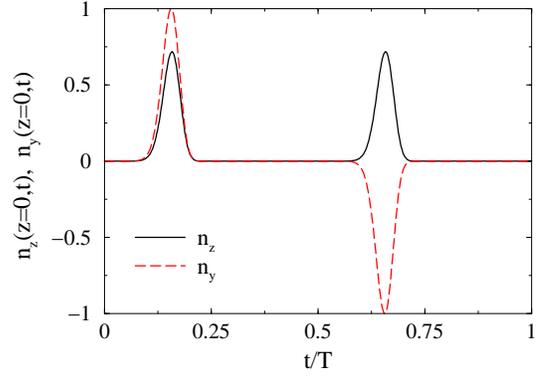}
\caption{(Color online) The director fields as a function of time at flexodomains threshold for a periodic excitation.
$\omega \tau_d = 0.05$, $\delta k = 0.3$ and $\xi_{-}=2$ (conductive symmetry).}
\label{fig:profnynz}
\end{figure}

While the onset of flexodomains in the presence of a dc-voltage is well described by the closed expression Eq.~(\ref{eq:bc_det}), the case of an ac-voltage requires in general numerical methods along the lines described in Sec.~\ref{sec:basic_eqns}.
We consider only low frequencies $\omega$ where $\omega \tau_d < 20$ which corresponds to frequencies $f=\omega/(2\pi)$ up to $20$~Hz for a cell of thickness $d =10$~$\mu$m filled with MBBA.
Note that we will sometimes use in the ac-case instead of the voltage amplitude, $U_0=E_0 d$, the {\em effective} (rms) voltage $U_0/\sqrt{2}$, which derives from the time average of $U^2(t)$.
As a warm-up we first consider the Freedericksz transition ($\mu \propto \epsilon_a >0$) at finite $\omega$, where $\bar{n}_y \equiv 0$ and $p =0$ in Eqs.~(\ref{eqn:nyz}).
In analogy to section~\ref{sec:dcflexo} we obtain the solution $\bar{n}_z(z,t) = \tilde{n}_z(t) \sin(z+\pi/2)$ with:
\begin{eqnarray}
\label{eq:nzom}
\tilde{n}_z(t) = \tilde{n}_z(0) 
\exp\left\{ -\int_0^t dt' [1 +\delta k -\mu u^2 \cos^2 (\omega t')] \right\} \;.
\end{eqnarray}
The solution $\tilde{n}_z(t)$ is periodic with period $T = 2 \pi /\omega$ when the integral in Eq.~(\ref{eq:nzom}) vanishes for the upper limit $t = T$.
Thus the Freedericksz transition voltage, $u_{Fr}(\omega)$, for $\omega \ne 0$ is obtained as:
\begin{eqnarray}
\label{eq:fredom}
\left[ u_{Fr}(\omega)/\sqrt{2} \right]^2 = (1+ \delta k)/\mu = u^2_F \;.
\end{eqnarray}
Obviously it is the (non-dimensionalized) {\it effective} (rms) critical voltage $u_{Fr}(\omega)/\sqrt(2)$, which continuously approaches in the limit $\omega \to 0$ the Freedericksz transition voltage $u_F$ (see Eq.~(\ref{eq:uF}) in the dc-case.
At onset for $u=u_{Fr}$ the solution Eq.~(\ref{eq:nzom}), normalized to its maximal value, has the following form
\begin{eqnarray}
\label{eq:nzom_c}
\tilde{n}_z(t) = 
\exp\left\{ -(1+\delta k) \frac{1 -\sin(2\omega t)}{2\omega} \right\} \;.
\end{eqnarray}
At small $\omega$ Eq.~(\ref{eq:nzom_c}) represents sudden burst-like director distortions similar to those shown in Fig.~\ref{fig:profnynz}.
The peaks of the $\tilde{n}_z(t)$ are located at $\omega t =\pi/4, 5\pi/4$, etc. and their width is proportional to $\sqrt{\omega}$.
Now we turn to the discussion of flexodomains where $n_y \ne 0$ and $p \ne 0$.
The threshold $u_c$ and the critical wavenumber $p_c$ for finite frequency $\omega$ are calculated as explained in Sec.~\ref{sec:basic_eqns}.
As a rule we get two types of solutions, one with the conductive, the other with the dielectric temporal symmetry.
In the following we use in general the material parameters of MBBA given in Eq.~(\ref{eq:mat}), where $k_{av}=5.43$, $\delta k = 0.2265$ [see Eq.~(\ref{eq:dk})] and $e_1-e_3=1.34$.
The dielectric torque is stabilizing, since $\epsilon_a=-0.53$.
According to our investigations the existence of flexodomains in the dc-case seems to be in general a necessary prerequisite for their existence in the ac-case.
In fact, at $\omega =0$ flexodomains solutions do not exist for standard MBBA, since according to Eq.~(\ref{eq:mu}) we have $\mu=-1.6027 < \mu_{min}(\delta k) = -0.815$; they have also not been observed in experiments. 
Thus we follow an idea in \cite{Thom:1989} and replace the difference $(e_1 -e_3)$ of the flexo coefficients in Eq.~(\ref{eqn:nyz}) by the product $\xi_{-} \cdot (e_1 -e_3)$ to study specifically the impact of the flexo torque.
Flexodomains then exist for $\xi_{-} > 1.402$.
Representative examples for the dependence of the critical voltage $u_c$ and the critical wavenumber $p_c$ on the ``flexo strength'' $\xi_{-}$ and on the ac-frequency $\omega$ (in units of $\tau_d^{-1}$) are shown in Fig.~\ref{fig:uc_e13mod}.
For $\xi_{-}=3$, corresponding to $\mu = -0.178$, the conductive symmetry prevails in the whole frequency range, while for $\xi_{-}=2$, corresponding to $\mu = -0.401$, a switch over from the conductive symmetry to the dielectric symmetry takes place at larger $\omega \tau_d$.
For both symmetry types the $u_c$ and $p_c$ curves rise monotonously as function of $\omega \tau_d$ at fixed $\xi_{-}$.
On the other hand, for the same symmetry type $u_c$ and $p_c$ are seen to decrease with increasing $\xi_{-}$, i.e., with growing magnitude of the flexo torque in analogy to the dc-case.
%

%%% Figure 7
%
\begin{figure}[ht]
\centering
\includegraphics[width=7.0cm]{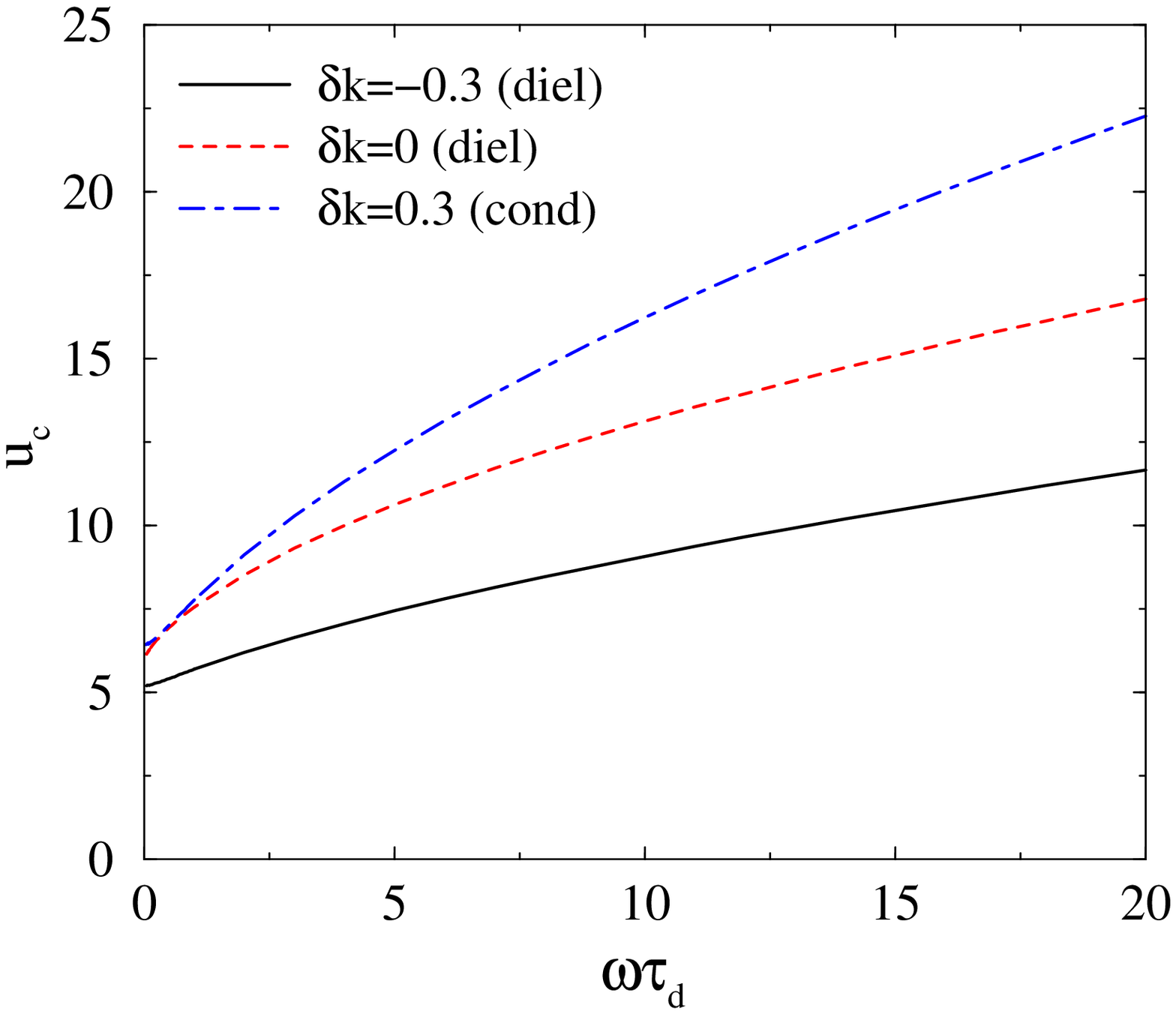}(a)
\includegraphics[width=7.0cm]{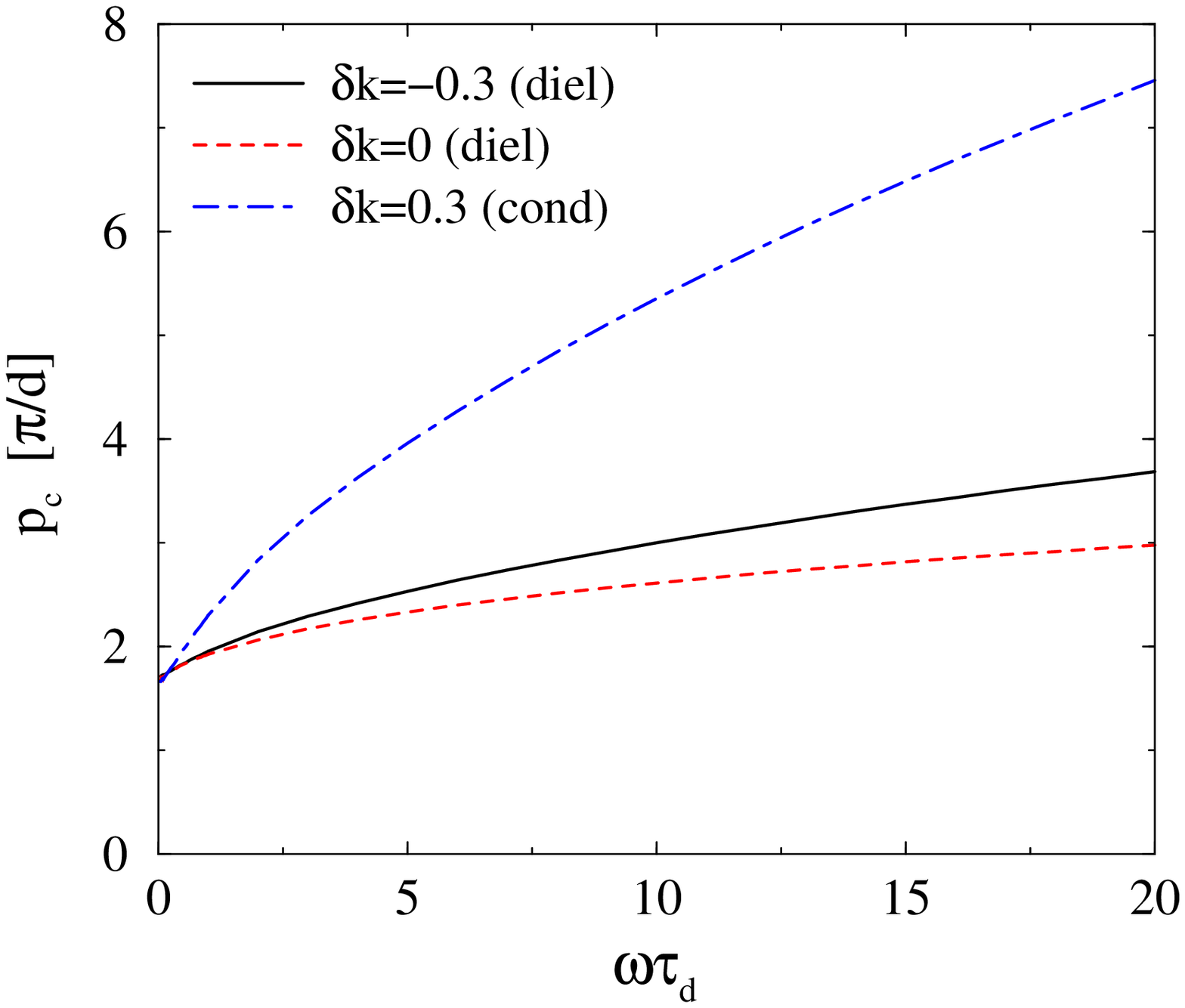}(b)
\caption{(Color online) Critical voltage $u_c(\omega)$ in dimensionless units (a) and critical wavenumber (b) of the flexodomains as function of $\omega \tau_d$ for different $\delta k$ and $\xi_{-}=2$.}
\label{fig:uc_dkmod}
\end{figure}

To study the dependence of the critical voltage $u_c(\omega)$ and the critical wavenumber $p_c(\omega)$ on the elastic constants, still MBBA parameters have been used, except a change of $\delta k$.
Furthermore we have chosen $\xi_{-}=2$ to favor the flexodomains against the homogeneous basic state.
The resulting data are shown in Fig.~\ref{fig:uc_dkmod}.
As before $u_c$ and $p_c$ increase monotonically with $\omega \tau_d$.
We are, however, unable to offer a simple explanation for the change of symmetry with $\delta k$: While the dielectric symmetry of the flexodomains solutions prevails for $\delta k = -0.3, \,0$, we find the conductive symmetry for $\delta k = 0.3$.
The switch over between the symmetries happens at $\delta k \approx 0.1$.
%

%%% Figure 8
%
In Fig.~\ref{fig:profnynz} we show an example of the director dynamics in the midplane ($z=0$) as a function of time at the low frequency $\omega \tau_d =0.05$.
Remarkable are the sudden burst-like director distortions, i.e., patterns would appear in the experiments only for very short time intervals.
The conductive time symmetry for $\delta k =0.3$ is reflected in the finite time average of $n_z(t)$; it vanishes for $n_y(t)$.
On the other hand, for $\delta k < 0.1$ the behavior of $n_z$ and $n_y$ with the dielectric time symmetry would be just opposite.
%

%%% Figure 9
%
Before we have concentrated on the MBBA-like materials with $\mu <0$.
In the case of $\mu >0$ we are faced with the competition of flexodomains with the Freedericksz transition as discussed in Sec.~\ref{sec:dcflexo} for dc-case.
The threshold voltage for the flexodomains increases monotonically with $\omega$ similar to the case of negative $\mu$ (see Figs.~\ref{fig:uc_e13mod}, \ref{fig:uc_dkmod}).
Since on the other hand $u_{Fr}(\omega)$ in Eq.~(\ref{eq:fredom}) does not depend on $\omega$, the Freedericksz transition will replace the flexodomains in any case above a certain finite $\omega$.
The frequency range where the flexodomains prevail will shrink with increasing $\mu$.
%

%%%
%
\subsubsection*{The limit $\omega \to 0$}
\begin{figure}[ht]
\centering
\includegraphics[width=7.0cm]{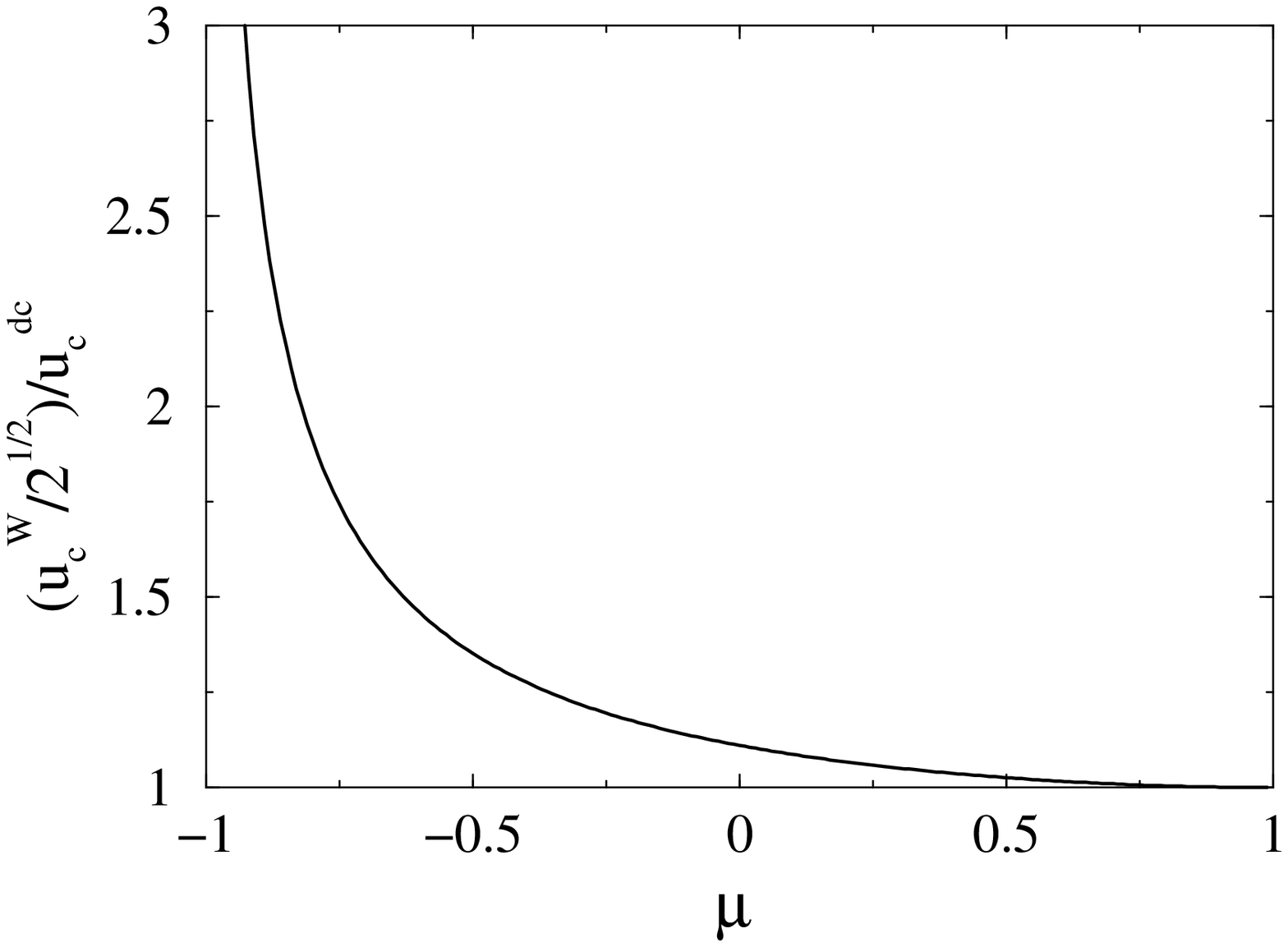}(a)
\includegraphics[width=7.0cm]{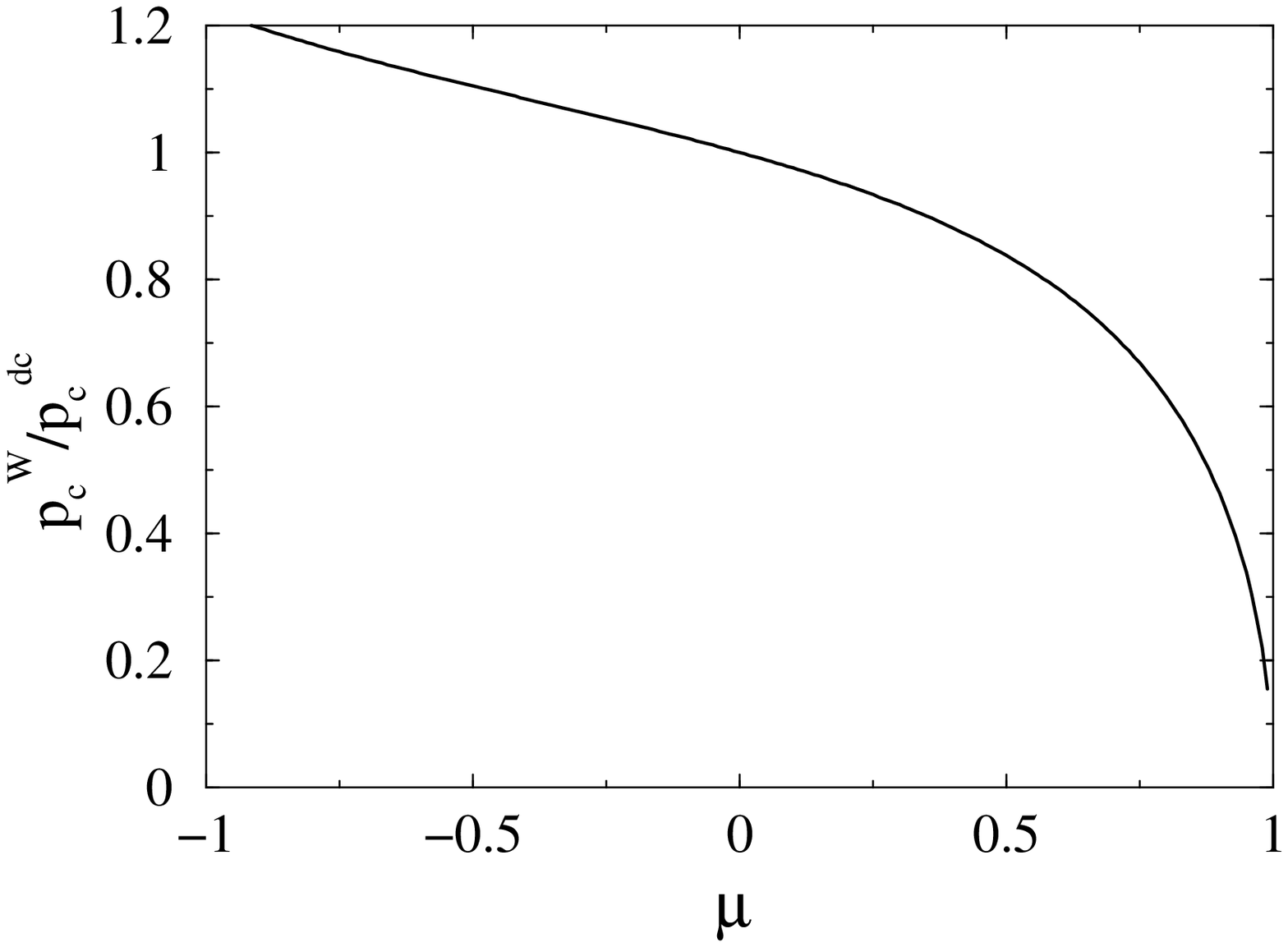}(b)
\caption{The critical effective voltage for flexodomains $u_c^{W}/\sqrt{2}$ (a) and the critical wavenumber $p_c^{W}$ (b) from Eq.~(\ref{eq:wkb}) in the limit $\omega \to 0$ normalized to the dc-values.}
\label{fig:ucpc_wkb_mu}
\end{figure}

The limit $\omega \to 0$ is obviously not smooth: In the dc-case both $n_z$ and $n_y$ have finite amplitudes at onset, while in the ac-case the time-average of one of the director components vanishes depending on the time symmetry.
An interesting question is, how this discontinuity is reflected in the critical voltage.
As a rule of the thumb it is typically assumed, that the effective (rms) voltages in the limit $\omega \to 0$ do not differ too much from critical voltages in the dc-case.
For the Freedericksz transition that difference is even zero [see Eq.~(\ref{eq:fredom})] and for electroconvection it is of the order of $10\%$ (see the Appendix of \cite{Bodenschatz:1988}).
In the case of flexodomains the corresponding differences are considerably larger in most cases.
As a representative example we discuss the $\omega \to 0$ limit of the $u_c$, $p_c$ curves in Fig.~\ref{fig:uc_e13mod} in comparison with the dc-case from Eq.~(\ref{eq:bc_det}).
Here MBBA material parameters have been used except that we allow for a modification of $e_1 -e_3$ via a factor $\xi_{-}$ introduced above.
At low frequencies the threshold curves both for $\xi_{-} = 2$ and $\xi_{-} = 3$ have conductive symmetry.
For $\xi_{-} = 2$ ($\mu = -0.401$) we find for instance $u_{eff}(0) \equiv u_c(\omega \to 0)/\sqrt{2} = 4.52$ compared to the dc-value $u_c(0) = 3.49$; for $\xi_{-} = 3$ ($\mu = -0.178$) the difference between $u_{eff}(0) = 2.83$ and $u_c(0) = 2.42$ is smaller.
The discontinuity of $u_{eff}(\omega)$ at $\omega = 0$ is associated with corresponding discontinuities of $p_c$: we find $p_c(\omega \to 0) = 1.66$ compared to $p_c(0) = 1.5$ for $\xi_{-} = 2$ and $p_c(\omega \to 0) = 1.25$ compared to $p_c(0) = 1.17$ when $\xi_{-} = 3$.
For the parameters chosen in Fig.~\ref{fig:uc_dkmod} we find comparable discontinuities as in Fig.~\ref{fig:uc_e13mod} for $\xi_{-} = 2$, where the symmetry of the director configuration (dielectric for $\delta k = -0.3, \,0$ and conductive for $\delta k = 0.3$) does not seem to play a significant role.
The discontinuities in the limit $\omega \to 0$ can be demonstrated most clearly in the special case $\delta k =0$.
Then the $z$-dependence of the director fields in Eqs.~(\ref{eqn:nyz}) is rigorously described by the ansatz $\{\bar{n}_y(z,t), \bar{n}_z(z,t)\} = \{\hat{n}_y(t), \hat{n}_z(t)\} \sin(z+\pi/2)$.
The transformation of the time variable $t \to t/\omega$ leads to: 
\begin{eqnarray}
\label{eqn:n1yfull}
\omega \partial_t \hat{n}_y &=& 
 -(p^2 + 1)\hat{n}_y + \sgn(e_1 -e_3) p u \cos(t) \hat{n}_z \;, 
\nonumber \\
\omega \partial_t \hat{n}_z &=& 
 -[p^2 + 1 - \mu u^2 \cos^2(t)]\hat{n}_z 
\nonumber \\
&& + \sgn(e_1 -e_3) p u \cos(t) \hat{n}_y \;.
\end{eqnarray}
From a mathematical point of view Eqs.~(\ref{eqn:n1yfull}) present an interesting dynamical system.
It is kind of ``singular'' since the prefactor of the time derivatives is proportional to $\omega$ and thus small in the limit $\omega \to 0$.
For the special case $\mu=0$ ($\epsilon_a =0$) the general solution of Eqs.~(\ref{eqn:n1yfull}) is given as:
\begin{eqnarray}
\label{ny_sol}
&& \hat{n}_y(t) = \exp\left[ -\frac{p^2+1}{\omega} t \right] 
\left\{ c_1 e^{\phi(t)} - c_2 e^{-\phi(t)} \right\} \;,
\nonumber\\
&& \hat{n}_z(t) = \exp\left[ -\frac{p^2+1}{\omega} t \right] 
\left\{ c_1 e^{\phi(t)} + c_2 e^{-\phi(t)} \right\} \;,
\nonumber \\
&& \text{with}\;
\phi(t) = \frac{ \sgn(e_1 -e_3) p u \sin(t)}{\omega} \;,
\end{eqnarray}
where $c_1$, $c_2$ denote integration constants.
Inspection of Eqs.~(\ref{ny_sol}) shows immediately, that one is unable to find values of $c_1$, $c_2$ to allow for periodic solutions. 
This result is a further hint, that the limit $\omega \to 0$ is not trivial, since in the dc-case according to Eqs.~(\ref{eq:pcpik}) flexodomains do exist at $\mu \propto \epsilon_a =0$.
The system Eqs.~(\ref{eqn:n1yfull}) for $\mu \ne 0$ has been in general investigated by the methods presented in Sec.~\ref{sec:basic_eqns}.
To study in particular the limit $\omega \to 0$ we have employed the approximation scheme presented in the Appendix of \cite{Bodenschatz:1988}, which is closely related to the familiar WKB approximation \cite{Bender} in quantum mechanics.
We use the ansatz $\{\hat{n}_y(t), \hat{n}_z(t)\} = \{\tilde{n}_y(t), \tilde{n}_z(t)\} \exp[S_0(t)/\omega]$ and neglect the derivatives of $\tilde{n}_y(t)$, $\tilde{n}_z(t)$.
The resulting coupled homogeneous equations for $\tilde{n}_y(t)$, $\tilde{n}_z(t)$ will have a nontrivial solution when the corresponding determinant vanishes.
In this way the following equation for $S_0(t)$ is obtained:
\begin{eqnarray}
\label{eq_wkb}
(\partial_t S_0)^2 + g_1(t) \partial_t S_0 + g_0(t) = 0 \;,
\end{eqnarray}
where
\begin{eqnarray}
\label{eq:g1_g0}
&& g_1(t) = 2 (p^2 + 1) - \mu u^2 \cos^2(t) \;,
\nonumber \\
&& g_0(t) = (p^2 + 1)^2 - [ p^2 + \mu(p^2 + 1) ] u^2 \cos^2(t) \;. \;\;
\end{eqnarray}
From Eq.~(\ref{eq_wkb}) we obtain:
\begin{eqnarray}
\label{eq:S0}
2 \partial_t S_0 = - g_1 + \sqrt{ g_1^2 - 4 g_0} \;.
\end{eqnarray}
In analogy to the discussion of the Freedericksz transition [see Eq.~(\ref{eq:nzom})] the solution is bounded when the average over one period ($2\pi$) of the right hand side of Eq.~(\ref{eq:S0}) vanishes.
The required time integration can be performed analytically and leads to the following implicit equation for the neutral curve $u_0(p) \equiv u_0(p,\omega \to 0)$
\begin{eqnarray}
\label{eq:wkb}
&& \pi [ 2 (p^2+1) - \mu u_0^2/2 ] |\mu| =
\nonumber \\ 
&& 2 \sqrt{p^2 \mu^2 u_0^2} + (4 p^2 + \mu^2 u_0^2)
\arcsin\left( \sqrt{\frac{\mu^2 u_0^2}{4 p^2 + \mu^2 u_0^2}} \right) \;.
\end{eqnarray}
This transcendental equation is solved numerically, which yields within the WKB approximation the critical voltage $u_c^{W}$ and the critical wavenumber $p_c^{W}$ of the flexodomains in the limit $\omega \to 0$.
The data agree perfectly with the exact solution of Eq.~(\ref{eqn:n1yfull}) on the basis of the monodromy matrix method in the limit $\omega \to 0$.
In Fig.~\ref{fig:ucpc_wkb_mu}(a) the ratio of dimensionless effective (rms) critical voltage $(u_c^{W}/\sqrt{2})$ and the dc-value is plotted as function of $\mu$.
While this ratio is near one for $-1/2 < \mu < 1$ it increases strongly when further decreasing $\mu$ towards $\mu = \mu_{min}(\delta k =0) =-1$.
The corresponding ratio of the critical wavenumber $p_c^{W}$ and the dc-values is shown in Fig.~\ref{fig:ucpc_wkb_mu}(b).
In comparison to the critical voltages one finds even larger deviations of this ratio from one in particular at larger positive $\mu$ where $p_c$ itself becomes small.
%

%%% Figure 10
%
\begin{figure}[ht]
\centering
\includegraphics[width=7.0cm]{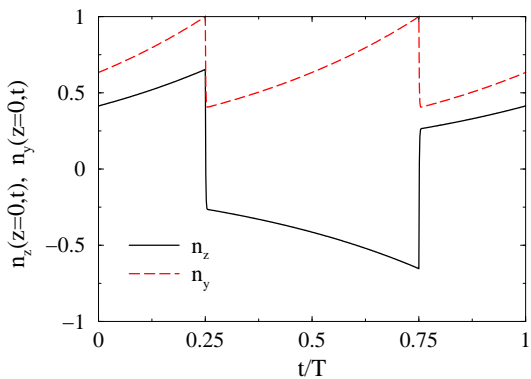}
\caption{(Color online) The director fields as function of time at flexodomains threshold for a square-wave excitation.
$\omega \tau_d = 0.05$, $\delta k =0$ and $\xi_{-}=2$ (dielectric symmetry).}
\label{fig:dk0_profnynz_sqr}
\end{figure}

Finally we would like to point out, that the transition to the dc-limit and the interesting time evolution of director perturbations characterized by bursts depend also on the wave-form of the exciting ac-voltage.
To demonstrate this we have considered a square-wave excitation which consists of an alternating sequence of constant voltages $\pm u$ on time intervals of the length $T/2$, i.e., $U(t)=U_0 \sgn[\cos(\omega t)]$.
Then Eqs.~(\ref{eqn:nyz}) can be solved quasi-analytically by joining continuously the analytical solutions on the parts with constant voltage, which leads to an implicit transcendental equation (``boundary determinant'') for the neutral curve $u_0(p)$.
Since for small $\omega \tau_d$ the values of $u_c$ and $p_c$ are practically determined by the long-time intervals of constant $u$ they are practically
identical to the dc-values.
Consequently, as demonstrated in Fig.~\ref{fig:dk0_profnynz_sqr}, the time variations of $n_z$, $n_y$ are rather smooth compared to the bursts observed with the harmonic ac-driving at low $\omega$.
%

%%% Figure 11
%
In the following section, we will study the alternative electroconvection (EC) solutions of the nemato-hydrodynamic equations, which compete with the flexodomains.
In general it will turn out, that only for small $\omega$ the flexodomains have the chance to prevail.
%

%%%
%
\section{Electroconvection}
\label{sec:ec}
\begin{figure}[H]
\centering
\includegraphics[width=7.0cm]{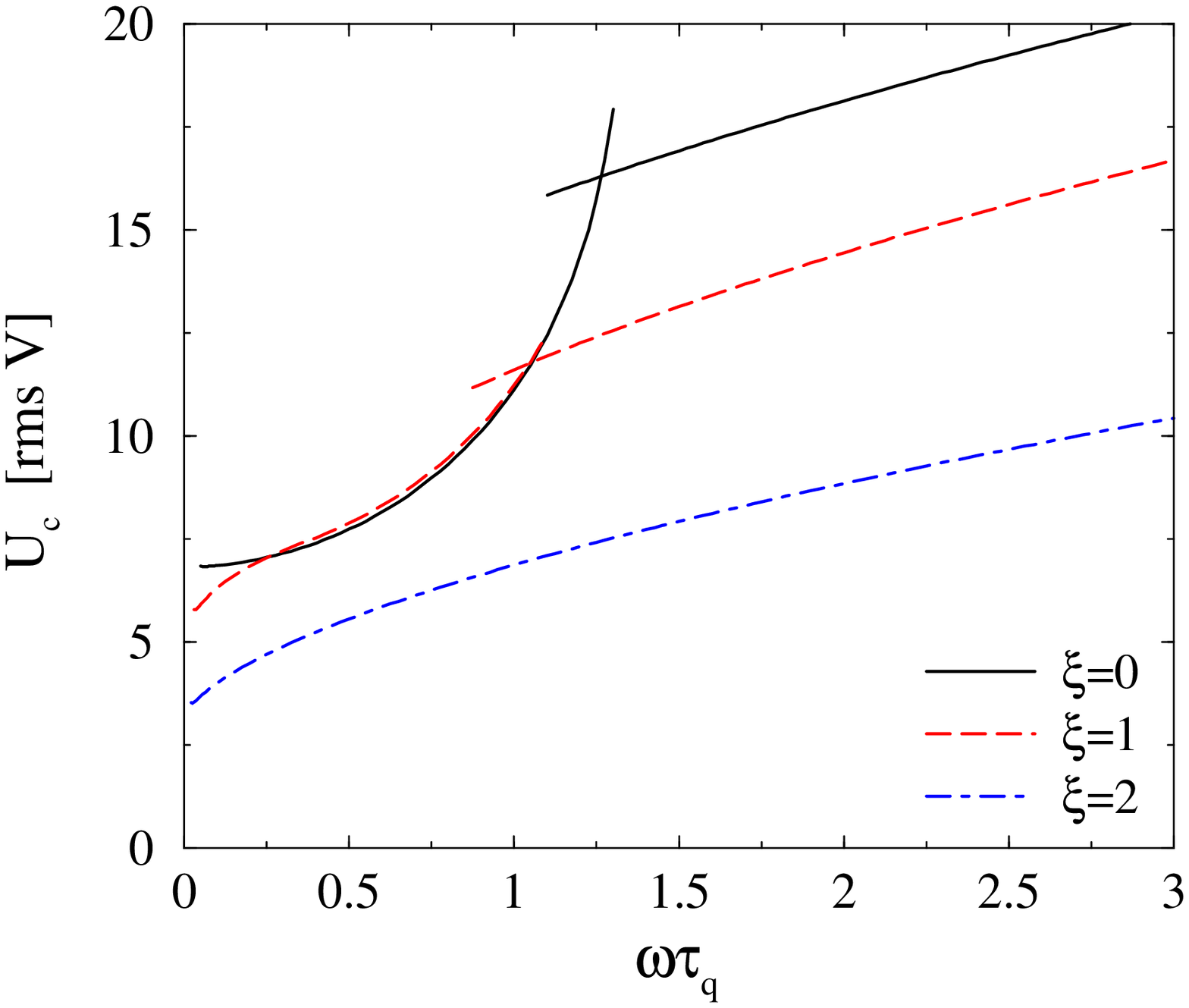}(a)
\includegraphics[width=7.0cm]{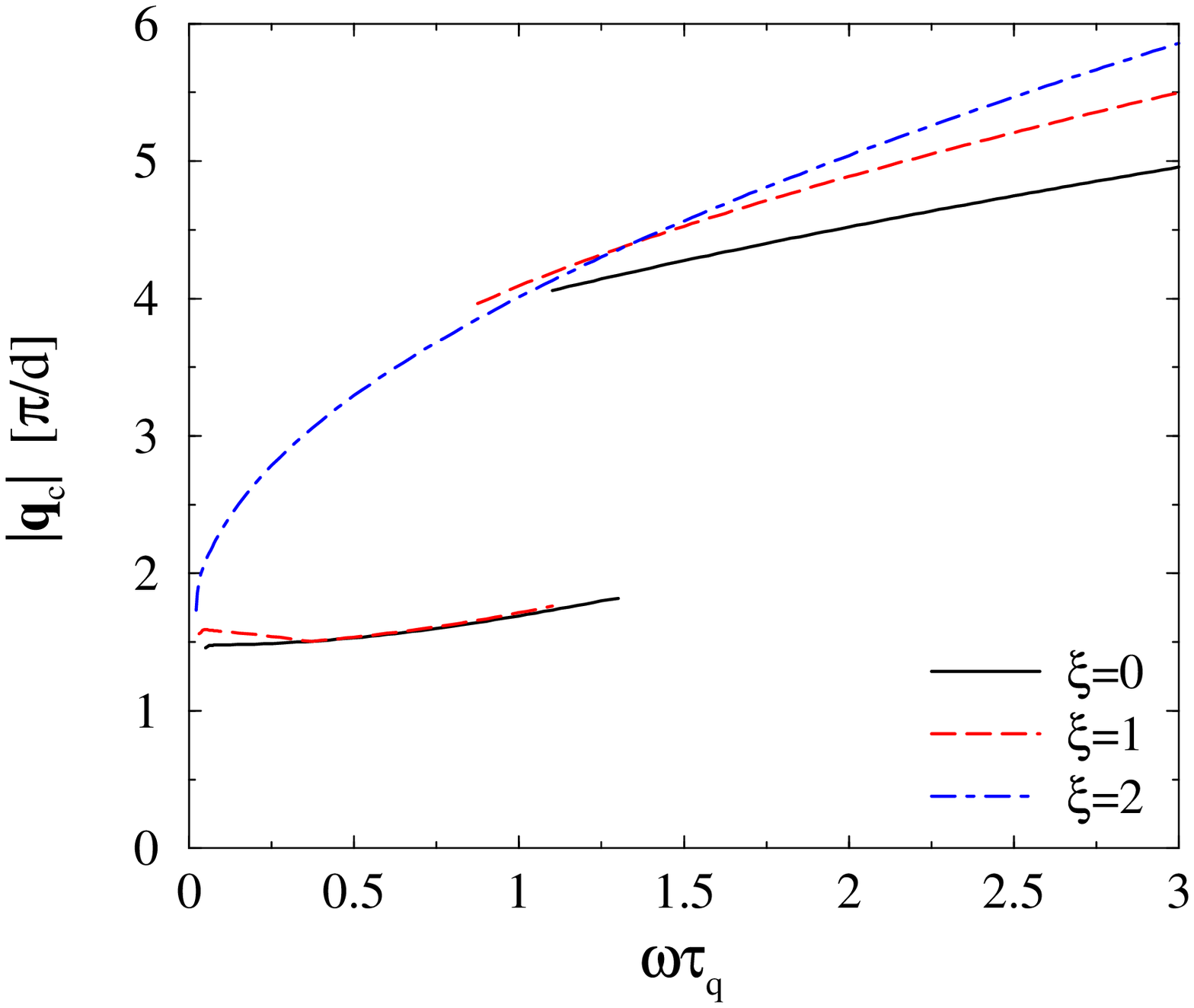}(b)
\includegraphics[width=7.0cm]{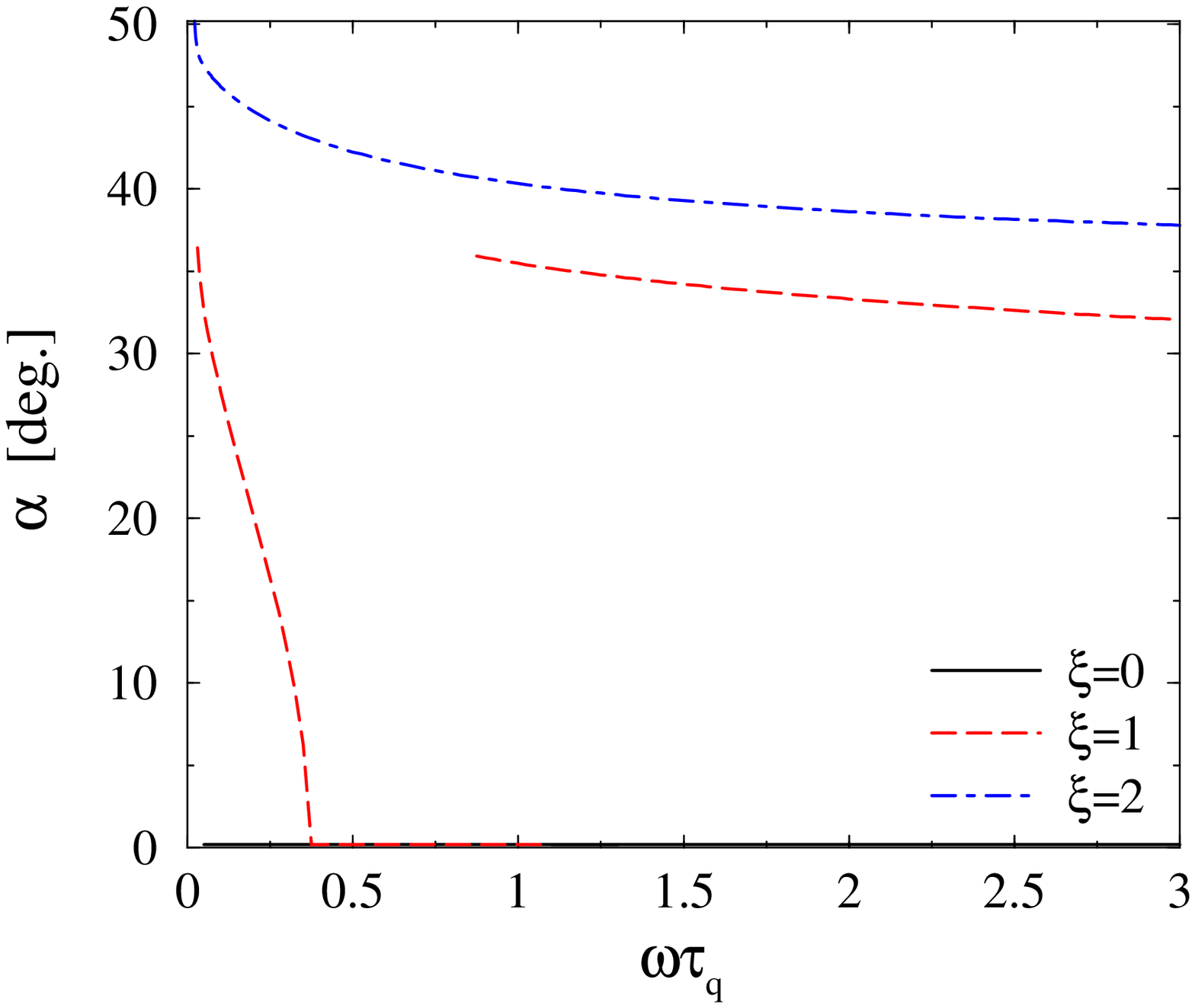}(c)
\caption{(Color online) Critical effective voltage $U_c$ of electroconvection (rms value in volt) (a), modulus of the critical wavevector $|{\bf q}_c|$ (b), and the angle $\alpha$ of the critical wavevector with the $x$-axis (c) as a function of frequency $\omega$ (in units of the charge relaxation time $\tau_q$) for different magnitudes of the flexo coefficients $\xi=\xi_{-}=\xi_{+}$ (for more details, see text).}
\label{fig:ec_wtq_e1e3}
\end{figure}

\begin{figure}[ht]
\centering
\includegraphics[width=7.0cm]{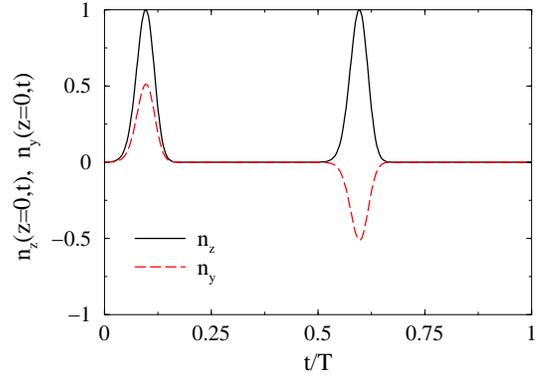}
\caption{(Color online) The director fields as a function of time at EC threshold for $\omega \tau_q =2.3 \times 10^{-3}$ ($\omega \tau_d =0.1$).}
\label{fig:nzny_t_ec}
\end{figure}

The salient elements of the positive feedback loop leading to the dissipative 
electro-hydrodynamic instability (EC), which are contained in the familiar nemato-hydrodynamic equations (see Sec.~\ref{sec:basic_eqns}), have been elucidated by Carr and Helfrich \cite{Carr:1969, Helfrich:1969}: A necessary condition is that the nematic is ``contaminated'' by a small amount of mobile ions, which results in a finite, though very small electrical conductivity of the order of $\sigma_0 = 10^{-8}$~($\Omega$~m)$^{-1}$.
Then spatial fluctuations of $\bm{n}$ in the presence of a nonzero $\bm{E}$ lead to an electric charge density $\rho_{el} = \bm{\nabla} \cdot \bm{D} + \rho_{fl}$.
The first contribution to $\rho_{el}$ is the standard one, which derives from the dielectric displacement $\bm{D} = \epsilon_0 \bm{\epsilon} \bm{E}$.
The latter ``nonstandard'' contribution, the flexo charge $\rho_{fl}$, is determined by the earlier introduced flexo polarization as follows: $\rho_{fl}= \bm \nabla \cdot \bm {P}_{fl}$.
Via the Coulomb force, $\rho_{el} \bm{E}$, in the (generalized) Navier-Stokes equation a material flow is driven, $\bm{v}$, which exerts an additional viscous torque on the director.
Under a favorable constellation of the material parameters the viscous torque reinforces the initial director distortion leading to EC.
For more details in particular a most recent review \cite{Buka:2006} might be useful, where systematically the sensitive influence of the sign of the dielectric anisotropy $\epsilon_a$, of the electric conductivity, $\sigma_a$, and of the basic director configuration on the patterns is discussed.
Exploiting the calculational scheme briefly discussed in section~\ref{sec:basic_eqns} we have precisely characterized the onset of EC.
While in flexodomains the time scale is set by the director relaxation time $\tau_d \propto d^2$ [see Eq.~(\ref{eq:mu})], in electroconvection the thickness independent charge relaxation time $\tau_q = \epsilon_0 \epsilon_{\perp}/\sigma_{\perp}$ plays an important role as well.
Thus we observe in EC a thickness dependence of the critical properties, which cannot be absorbed by a suitable frequency rescaling.
For definiteness we concentrate in the following on a nematic layer of thickness $d=10$~$\mu$m and use MBBA parameters (if not otherwise stated), where $\tau_d = 0.2$~s and $\tau_q = 4.65 \times 10^{-3}$~s. 
The impact of the flexo charge $\rho_{fl}$ on EC is determined by the sum $(e_1 + e_3)$ of the flexo coefficients.
In analogy to the factor $\xi_{-}$ introduced in section~\ref{sec:pikin} to tune the strength of the flexo torque, we replace here in addition $(e_1 + e_3)$ by $\xi_{+} \cdot (e_1 + e_3)$ to allow for a selective modification of $\rho_{fl}$.
To facilitate a direct comparison with experiments we will refer to effective (rms) critical voltages in physical units everywhere in this section. 
In Fig.~\ref{fig:ec_wtq_e1e3} we present representative critical EC-data for MBBA parameters \cite{remark} as function of $\omega \tau_q$ for different $\xi =\xi_{-} = \xi_{+}$.
In general $U_c$ and $|\bm q_c|$ [see Figs.~\ref{fig:ec_wtq_e1e3}(a), (b)] increase monotonically with increasing $\omega$.
While the two curves are smooth for $\xi = \xi_{-} = \xi_{+} =2$, where the dielectric symmetry prevails, we observe discontinuities for the two smaller $\xi$.
Here the conductive branch at small $\omega$ is replaced by the dielectric one at a crossover frequency $\omega_c$ where $\omega_c \tau_q \approx 1.2$.
The obliqueness of the rolls, as shown in Fig.~\ref{fig:ec_wtq_e1e3}(c), is measured by the angle $\alpha$ between $\bm{q}_c$ and the preferred director orientation in the basic planar configuration (our $x$-axis).
In the absence of flexo polarization ($\xi =0$) we find $\alpha \equiv 0$ for all $\omega$.
When $\xi_{-} = \xi_{+} =1$ the obliqueness of the rolls in the conductive regime vanishes continuously at the ``Lifshitz point'' $\omega_L \tau_q \approx 0.35$ and becomes finite again in the dielectric regime.
At the largest $\xi = 2$ the angle $\alpha \approx 40^\circ$ is finite for all $\omega$.
Note that recently for other nematic material a ``Lifshitz point'' in the dielectric regime of EC has been identified as well \cite{May:2008}.
In analogy to the flexodomains the limit $\omega \to 0$ for the critical voltage in EC is not smooth.
However, the dc-critical voltage is much better approximated by the limit $\omega \to 0 $ of the effective voltage $U_c(\omega)$ than for the flexodomains.
This finding can be explained within a WKB approximation \cite{Bodenschatz:1988}, where the relative corrections are indeed small of the order of $\tau_q/\tau_d$ (up to some $q$-dependent factors).
EC shares, however, with the flexodomains the typical spiky time evolution of the director components at very low $\omega$, as comparison of Fig. \ref{fig:profnynz} with Fig.~\ref{fig:nzny_t_ec} shows.
%

%%% Figure 12
% 
%

%%% Figure 13
% 
\begin{figure}[ht]
\centering
\includegraphics[width=7.0cm]{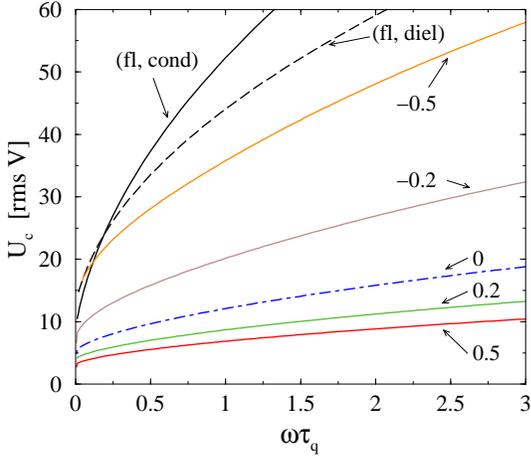}
\caption{(Color online) Critical voltage $U_c$ (rms value in volt) for EC as a function of frequency $\omega$ (in units of the charge relaxation time $\tau_q$) for five different values of $\sigma_a/\sigma_{\perp}$ between $0.5$ and $-0.5$ and for $\xi_{-} = \xi_{+} = 2$.
The corresponding critical voltages for the flexodomains (conductive symmetry at small $\omega$ and dielectric symmetry at larger $\omega$) are included as well.}
\label{fig:ucsa10}
\end{figure}

Besides the flexo coefficients a further crucial parameter is the anisotropy $\sigma_a$ of the electric conductivity, since the Carr-Helfrich charge separation mechanism depends strongly on the magnitude and the sign of $\sigma_a/\sigma_{\perp}$.
This dependence had not been investigated in detail for many years in the past, since in the materials used at that time $\sigma_a/\sigma_{\perp}$ would only vary between $0.3$ and $0.7$.
Recently, however, materials have been found, where by decreasing the temperature $\sigma_a$ decreases as well; it passes even zero and becomes negative.
Thus it has now become important and attractive to study the $\sigma_a$ dependence in more detail.
This has been done partially in \cite{Krekhov:2008}, where $U_c(\omega)$ has been observed to move up in general with decreasing $\sigma_a$.
To study the details we have taken MBBA parameters as before except that $\sigma_a/\sigma_{\perp}$ was allowed to vary.
Furthermore, to explore the competition with flexodomains, which do not exist for standard MBBA (see Sec.~\ref{sec:acflexo}) the flexo coefficients have to be increased as well.
We have centered on the case $\xi = \xi_{-} = \xi_{+} = 2$, which has been shown already in Fig.~\ref{fig:ec_wtq_e1e3} for EC and in Fig.~\ref{fig:uc_e13mod} for flexodomains.
The EC threshold, which is characterized by the dielectric symmetry, increases strongly with decreasing $\sigma_a/\sigma_{\perp}$ for all $\omega$ as documented in Fig.~\ref{fig:ucsa10}.
Only for $\sigma_a/\sigma_{\perp}= -0.5$ and for very small frequencies up to $\omega \tau_q \approx 0.1$ (where $\omega \tau_d \approx 3.6$) flexodomains with conductive symmetry prevail.
For the case $\xi_{-} = \xi_{+} = 1$ one finds the same behavior of $U_c(\omega)$ when changing $\sigma_a$, while the flexodomains do not exist.
% 

%%% Figure 14
% 
\begin{figure}[ht]
\centering
\includegraphics[width=7.0cm]{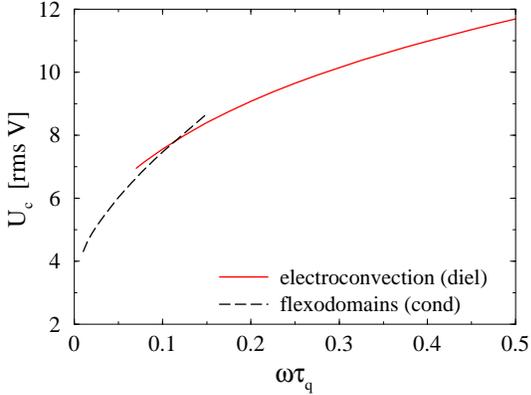}
\caption{(Color online) Critical effective voltage $U_c$ (rms value in volt) as a function of frequency $\omega$ (in units of the charge relaxation time $\tau_q$) for flexodomains and EC patterns.
Material parameters of MBBA except $\sigma_a/\sigma_{\perp} =0.2$, $\xi_{-} =3$, and $\xi_{+} = 0.25$.}
\label{fig:ecpik}
\end{figure}

To find material combinations, where flexodomains have a chance to dominate the pattern
forming instability for standard and not too small values of $\sigma_a$, one has
obviously to chose different values of $\xi_{-}$ and $\xi_{+}$.
One needs a relatively large $\xi_{-}$ to increase the flexo torques, which are responsible
for the existence of flexodomains.
Conversely, $\xi_{+}$ should be fairly small to suppress the positive influence of the flexo charge on EC.
Since systematic experimental studies, also including different nematic materials are missing so far, extended parameter studies seem to be futile at the moment.
So we present only one example with $\xi_{-} =3$, $\xi_{+} = 0.25$ and $\sigma_a/\sigma_{\perp} =0.2$ where flexodomains (with conductive symmetry) indeed prevail EC at low frequencies as shown in Fig.~\ref{fig:ecpik}.
%

%%% Figure 15
%
%

%%% Figure 16
%
\begin{figure}[H]
\centering
\includegraphics[width=7.0cm]{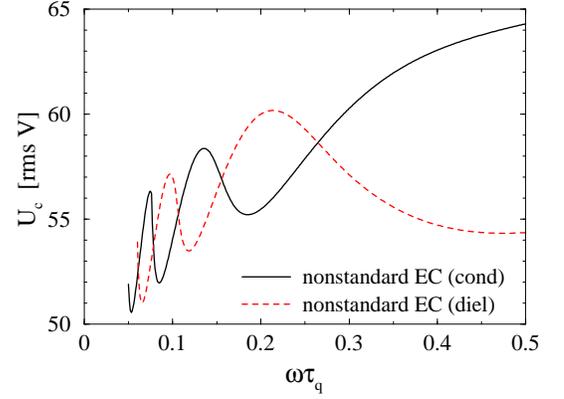}
\caption{(Color online) Critical effective voltage $U_c$ of electroconvection (rms value in volt) as a function of frequency $\omega$ (in units of the charge relaxation time $\tau_q$) for MBBA material parameters ($\xi_{+} = \xi_{-} =1$) except $\sigma_a/\sigma_{\perp}=-0.05$.}
\label{fig:ecosc}
\end{figure}
Finally, we address briefly the intriguing limiting behavior of EC at low frequencies, when $\xi_{-}$ is too small for the existence of flexodomains at $\omega =0$.
As demonstrated in Fig.~\ref{fig:ecosc} we predict in this situation a persistent switching between the conductive and the dielectric solutions for $\sigma_a/\sigma_{\perp}=-0.05$.
%

%%%
%
\section{Conclusions}
\label{conclusions}
In this paper we have investigated pattern forming instabilities in nematic liquid crystals originally in the basic planar configuration, which are driven out of equilibrium by an electric dc or ac voltage applied perpendicular to the layer.
In the basic state the director field is homogeneous over the nematic layer and a preferred direction is singled out in the layer plane.
In particular we have concentrated on the impact of the flexo effects, which generically exist, when the director varies in space.
They results in on one hand to the flexo torque which may lead at sufficiently large voltage amplitude to the flexodomains.
These are spatially periodic in the layer plane and can be visualized in experiments as stripes parallel to the preferred planar direction.
In this paper the bifurcation scenario to flexodomains is for the first time comprehensively analyzed for applied dc and ac voltages. 
We have also demonstrated that the well-established splay-twist Freedericksz transition in the absence of flexo effects can be described in this framework. 
Particular focus was on the intriguing phenomena in the limit of vanishing ac-frequency $\omega$ of the driving voltage.
One finds for instance periodic-in-time, burst-like excursion of the director from the
planar basic state.
Furthermore we have investigated the competition between the flexodomains and the familiar EC convective rolls.
The latter are visualized as stripes which include a finite angle typically larger than $45^{\circ}$ with the preferred axis.
In general EC prevails, except at very low frequencies where the flexodomains may trigger the destabilization of the homogeneous basic state.
Like the flexodomains the EC pattern appear also periodically in time as sudden bursts, while the system remains otherwise in the structureless basic planar state.
We expect that our theoretical analysis will motivate further investigations of electrically driven pattern formation in nematics with special focus on the low-frequency regime.
Though in particular the sudden, periodic-in-time ''blow-up'' of flexodomains and EC patterns as well a crossover between them have been clearly seen in recent experiments \cite{May:2008}, a quantitative comparison with the theory is far from trivial.
On one hand one is faced with considerable uncertainties in the material parameters even for a nematic like MBBA.
Moreover, one has to realize that the nematic layer is confined by metallically coated glass plates with a thin polymeric alignment film on top, which themselves may have fairly complicated electric properties.
Thus the whole system is represented as an equivalent circuit diagram (see, e.g., \cite{Ohe:1994, Seiberle:1994}) where the total voltage applied to the cell is different from ``theoretical'' results like $U_c$, which represent only the voltage drop over the nematic layer.
The necessary corrections to $U_c$ are not easy to model and would also require systematic measurements on the empty cell.
Finally the coexistence between flexodomains and EC pattern in the nonlinear regime when their linear thresholds are comparable has to be investigated.
For instance such a scenario has indeed been observed in hybrid aligned nematic cell \cite{Delev:2001} in the presence of a dc-voltage.
%

%%%
%
\begin{acknowledgments}
Financial support by the Deutsche Forschungsgemeinschaft Grant FOR-608 and the Hungarian Scientific Research Fund OTKA-K81250 are gratefully acknowledged.
\end{acknowledgments}
%

%%%
%
\appendix
\section{Flexodomains driven by a dc-voltage}
In this Appendix we describe briefly the derivation of the implicit equation (\ref{eq:bc_det}) which yields the neutral curve $u_0(p)$.
We use the ansatz $\{ \bar{n}_z(z), \bar{n}_y(z) \} = \{ \hat{n}_z, \hat{n}_y \} e^{\lambda z}$ in Eq.~(\ref{eqn:nyz}) and set $\sigma =0$ from the beginning.
Thus we arrive at a linear homogeneous $2 \times 2$ system for $\hat{n}_{y}$ $\hat{n}_{z}$ which leads to the following secular equation for $\lambda$:
\begin{eqnarray}
\label{eq:lambda}
&& \left[ p^2 (1 + \delta k) - (1-\delta k) \lambda^2 \right]
   \left[ p^2 (1 - \delta k) - \mu u^2 - (1+\delta k) \lambda^2 \right] 
\nonumber \\
&& -p^2(\sgn^2(e_1-e_3) u^2 - 4{\delta k}^2 \lambda^2) = 0 \;. \qquad\qquad\qquad\qquad
\end{eqnarray}
From Eq.~(\ref{eq:lambda}) we obtain the two eigenvalues $\lambda^2 = \lambda_1^2, -\lambda_2^2$, where $\lambda_1^2, \lambda_2^2 >0$, which read as follows:
\begin{eqnarray}
\label{l1l2}
&& \lambda_1^2 = p^2 - \frac{\mu u^2}{2(1+\delta k)} + S \;, \;\;
%\nonumber \\
   \lambda_2^2 = -p^2 + \frac{\mu u^2}{2(1+\delta k)} + S \;,
\nonumber \\
&& S =  
\left[ \frac{\sgn^2(e_1-e_3) + 2 \delta k \mu}{1-\delta k^2} p^2 u^2 
+ \left( \frac{\mu u^2}{2(1+\delta k)} \right)^2 \right]^{1/2} \;,
\end{eqnarray}
The eigenvector of the linear system is chosen as:
\begin{eqnarray}
\label{eq:eigv}
 \left\{ \hat{n}_z (\lambda), \hat{n}_y (\lambda) \right\} =
 \left\{ 1, p\frac{\sgn(e_1-e_3) u -2 \delta k \lambda}
                  {p^2 (1+\delta k) - (1-\delta k) \lambda^2} \right\} \;.
\end{eqnarray}
In line with the standard procedure we express the general solution as a linear combination of the eigenmodes (\ref{eq:eigv})
\begin{eqnarray} 
\label{eq:ansatz}
&& \bar{n}_z(z) = a_1 e^{\lambda_1 z } + a_2 
e^{-\lambda_1 z} + a_3 e^{i \lambda_2 z} + a_4 e^{-i \lambda_2 z} \;,
\nonumber \\
&& \bar{n}_y(z) = a_1 \hat{n}_y(\lambda_1) e^{\lambda_1 z } + a_2 \hat{n}_y(-\lambda_1) 
e^{-\lambda_1 z} 
\nonumber \\
&&\qquad\quad +  a_3 \hat{n}_y(i \lambda_2)e^{i \lambda_2 z}+ a_4 \hat{n}_y(-i  \lambda_2) e^{-i \lambda_2 z} \;,
\end{eqnarray}
to fulfill the boundary conditions $\bar{n}_z(\pm \pi/2) = \bar{n}_y(\pm \pi/2) =0$.
A nontrivial solution exists, when the determinant of resulting set of four linear homogeneous equations for the $a_i$ vanishes.
The resulting implicit equation for the neutral curve has been already given before in Eq.~(\ref{eq:bc_det}).
It contains the $\lambda_i$ (\ref{l1l2}) and coefficients $A_i$, which are given as follows:
\begin{eqnarray}
\label{eq:coefA}
&& A_1 = \left[ \left( f_1^2 \lambda_2^2 - f_2^2 \lambda_1^2 \right) C^2 
+ \left( \lambda_1^2 + \lambda_2^2 \right)^2 D^2 \right] \;,
\nonumber \\
&& A_2 = 2 f_1 f_2 C^2 \;,
\end{eqnarray}
with 
\begin{eqnarray}
\label{eq:defABC}
&& f_1 = \lambda_1^2 - B \;,\;\; f_2 = \lambda_2^2 + B \;, \;\;
   B = p^2 \frac{1+ \delta k }{1 - \delta k } \;,
\nonumber \\
&& C = \frac{2 \delta k }{\sqrt{1-\delta k^2}} \;,\;\;
   D = \frac{\sgn(e_1 -e_3)  u}{\sqrt{1-\delta k^2}} \;.
\end{eqnarray}
{\sc Maple} and {\sc Mathematica} have been very useful to perform and to validate all our calculations.
The Freedericksz transition in the absence of flexo effects ($e_1-e_3=0$) is formally covered by Eq.~(\ref{eq:bc_det}) as well.
First of all the voltage $u$ appears always in the combination $(\mu u^2)$ where according to Eq.~(\ref{eq:mu}) the factors $(e_1-e_3)$ cancel.
Furthermore we have $D \equiv 0$ [see Eq.~(\ref{eq:defABC})] in this limit.
The critical $\delta k_{ST}$ at which the splay-twist Freedericksz transition starts to prevail is obviously determined by the conditions $\partial^2_{p} u_0(p =0) = \partial^4_{p} u_0(p =0) = 0$ at $u_0(p)/u_F =1$.
This condition is exploited by expanding Eq.~(\ref{eq:bc_det}) to order $p^4$ and we arrive in fact at an expression for $\delta k_{ST}$ which is identical to $\delta k_c$ in Eq.~(\ref{eq:deltkc}).
Note that in \cite{Lonberg:1985} instead of exact value for $\delta k_c = 0.534624$ the approximation ${\delta k}_c \approx 0.53$ was given.
On the background of the exact textbook analysis presented in this Appendix it is easy to demonstrate, why the authors of \cite{Hinov:2009, Marinov:2010} came to wrong results.
Starting from the basic equations for the flexodomains in dc-case [equivalent to Eqs.~(\ref{eqn:nyz})] at first a set of two coupled ODE's that contain only second order $z$-derivatives has been obtained (see Eq.~(12) in \cite{Marinov:2010}).
Unfortunately the second step, namely diagonalizing the ODE's by a ``unitary'' transformation $\hat{\bm{V}}$ (see Eq.~(14) in \cite{Marinov:2010}), does not work since $\hat{\bm{V}}$ explicitly depends on $z$.
Thus the main conclusion in \cite{Marinov:2010} that the eigenmodes associated with the eigenvalues $P$ and $Q$ (see Eq.~(20) in \cite{Marinov:2010}) would decouple is untenable.
This applies also to the resulting simple relation $\tan(i Q d) =0$ (see Eq.~(22) in \cite{Marinov:2010}) which would replace our correct Eq.~(\ref{eq:bc_det}) for the neutral curve $u_0(p)$.
Note, that the same erroneous procedure has appeared already in a previous investigation \cite{Romanov:1997}, cited also in \cite{Hinov:2009, Marinov:2010}.
%

%%%
%

%


\begin{thebibliography}{99}
\bibitem{remark}
The flexo coefficients $e_1$, $e_2$ for MBBA are taken from \cite{Thom:1989}.
The other material parameters can be found for instance as the set MBBA~I in the Appendix of \cite{Bodenschatz:1988}.
\end{thebibliography}
\end{document}